\documentclass[12pt]{article}
\pdfoutput=1
\usepackage{jheppub}
\usepackage{epsfig}
\usepackage{amsmath}
\usepackage{amssymb}
\usepackage{amsfonts}
\usepackage{amsxtra}
\usepackage{amsthm}
\usepackage{mathrsfs}
\usepackage{makeidx}
\usepackage{graphics}
\usepackage{dsfont}
\usepackage{mathtools}
\usepackage{graphicx}
\usepackage{placeins}
\usepackage{bm}
\usepackage[capitalise]{cleveref}
\usepackage{empheq}
\usepackage{colortbl}
\usepackage{xcolor}
\usepackage{enumerate}
\usepackage{titlesec}
\usepackage{longtable}
\usepackage{hhline}
\usepackage{float}
\usepackage{color}
\usepackage{tikz}
\usepackage{xfrac}
\usepackage{footnote}
\usepackage{rotating}
\usepackage{lscape}
\usepackage{makecell}
\usepackage{environ}
\usepackage{tabularx}
\usepackage{subfiles}
\usepackage[export]{adjustbox}
\usepackage{ytableau}
\usepackage{tikz-3dplot}
\usepackage{slashed}
\usepackage{pifont}
\usepackage{multirow}
\usepackage{mdframed}
\usepackage{bbm}
\usepackage[normalem]{ulem}
\usepackage{enumitem} 
\usepackage{fancybox}
\usepackage[style=numeric-comp,sorting=none,maxbibnames=99,giveninits=true,backend=bibtex,backref=true]{biblatex}
\DeclareFieldFormat{doilink}{
\iffieldundef{doi}{#1}
{\href{http://dx.doi.org/\thefield{doi}}{#1}}}

\DeclareBibliographyDriver{article}{%
  \usebibmacro{bibindex}%
  \usebibmacro{begentry}%
  \usebibmacro{author/translator+others}%
  \setunit{\labelnamepunct}\newblock
  \usebibmacro{title}%
  \newunit
  \printlist{language}%
  \newunit\newblock
  \usebibmacro{byauthor}%
  \newunit\newblock
  \usebibmacro{bytranslator+others}%
  \newunit\newblock
  \printfield{version}%
  \newunit\newblock
  \usebibmacro{in:}%
  \href{http://dx.doi.org/\thefield{doi}}{%
  \usebibmacro{journal+issuetitle}%
  \newunit\newblock
  \usebibmacro{byeditor+others}%
  \newunit\newblock
  \usebibmacro{note+pages}}%
  \newunit\newblock
  \iftoggle{bbx:isbn}
   {\printfield{issn}}
   {}%
  \newunit\newblock
  \usebibmacro{doi+eprint+url}%
  \newunit\newblock
  \usebibmacro{addendum+pubstate}%
  \newunit\newblock
  \usebibmacro{pageref}%
  \usebibmacro{finentry}
  }

\DeclareFieldFormat{url}{\mkbibacro{URL}\addcolon\space\href{#1}{Link}}

\DefineBibliographyStrings{english}{%
  backrefpage = {Cited on page},
  backrefpages = {Cited on pages},
}

\usetikzlibrary{positioning,trees,decorations.pathmorphing,decorations.markings,decorations.pathreplacing,calc,shapes,patterns,arrows,chains,arrows.meta,fit,fadings,decorations.markings,graphs,graphs.standard,quotes,plotmarks}

\DeclareMathOperator{\U}{U}
\DeclareMathOperator{\SU}{SU}
\DeclareMathOperator{\SO}{SO}
\DeclareMathOperator{\SL}{SL}

\DeclareMathOperator{\USp}{USp}

\newcommand{\cN}{\mathcal{N}}

\newcommand{\coma}{\, , \quad}
\newcommand{\fstop}{\, .}

\def\ZZ{{\mathbb{Z}}}

\theoremstyle{plain}

\theoremstyle{definition}

\newtheorem{claim}{Claim}

\let\oldFootnote\footnote
\newcommand\nextToken\relax

\renewcommand\footnote[1]{%
    \oldFootnote{#1}\futurelet\nextToken\isFootnote}

\newcommand\isFootnote{%
    \ifx\footnote\nextToken\textsuperscript{,}\fi}

\tikzset{cross/.style={cross out, draw=black, fill=none, minimum size=2*(#1-\pgflinewidth), inner sep=0pt, outer sep=0pt}, cross/.default={2pt}}

\usetikzlibrary{positioning}
\usetikzlibrary{chains}
\usetikzlibrary{arrows, arrows.meta ,fit,decorations.pathreplacing}
\tikzstyle{every picture}+=[remember picture]
\tikzstyle{na} = [baseline]
\tikzstyle{ligne}=[draw, thick]

\usetikzlibrary{arrows, decorations.markings, calc, fadings, decorations.pathreplacing, patterns, decorations.pathmorphing, positioning}
\tikzset{>={Latex[width=1.5mm,length=1.5mm]}}
\tikzset{bd/.style={circle, draw=black, inner sep=0pt, fill=black, minimum size=1.2mm}}
\tikzset{bld/.style={circle, draw=blue, inner sep=0pt, fill=blue, minimum size=1.2mm}}
\tikzset{wd/.style={circle, draw=black, inner sep=0pt, fill=white, minimum size=1.2mm}}
\tikzset{rd/.style={circle, draw=red, inner sep=0pt, fill=red, minimum size=.9mm}}
\tikzset{wrd/.style={circle, draw=red, inner sep=0pt, fill=white, minimum size=.9mm}}
\usetikzlibrary{graphs,graphs.standard,quotes}

\def\node#1#2{\overset{#1}{\underset{#2}{{\color{gray} \bullet}}}}

\def\node#1#2{\overset{#1}{\underset{#2}{\circ}}}

\tikzstyle{every picture}+=[remember picture]
\tikzstyle{na} = [baseline=-.5ex]

\newcommand{\ie}{i.e. }

\numberwithin{equation}{section}
\newcommand{\bes}[1]{\begin{equation} \begin{split} #1\end{split} \end{equation}}

\newcommand{\be}{\begin{equation}} \newcommand{\ee}{\end{equation}}
\newcommand{\bea}{\begin{equation} \begin{aligned}} \newcommand{\eea}{\end{aligned} \end{equation}}

\def\tilde{\widetilde}

\def\rt2{\sqrt{2}}

\def\mod{{\rm mod}}

\def\CA{{\cal A}}
\def\CB{{\cal B}}

\def\CM{{\cal M}}
\def\CN{{\cal N}}

\def\CP{{\cal P}}

\def\CS{{\cal S}}
\def\CT{{\cal T}}

\def\1{{\ds 1}}

\newcommand{\bZ}{\mathbb{Z}}
\newcommand{\fg}{\mathfrak{g}}

\def\SO{\mathrm{SO}}

\def\SU{\mathrm{SU}}
\def\SL{\mathrm{SL}}

\def\su{\mathfrak{su}}

\def\so{\mathfrak{so}}

\def\repa{\raise4pt\hbox{$\square$}\mkern-14mu\raise-4pt\hbox{$\square$}}
\def\repab{\overline{\raise4pt\hbox{$\square$}\mkern-14mu\raise-4pt\hbox{$\square$}\mkern-1mu}}

\def\smileface{\ensuremath{\hbox{\large$\bigcirc$}\mkern-15mu\raise-1pt\hbox{\scriptsize$\smallsmile$}%
\mkern-10mu\raise4pt\hbox{..}\mkern4mu}}
\def\frownface{\ensuremath{\hbox{\large$\bigcirc$}\mkern-15mu\raise-1pt\hbox{\scriptsize$\smallfrown$}%
\mkern-10mu\raise4pt\hbox{..}\mkern4mu}}

\newcommand{\ba}{\begin{array}}
\newcommand{\ea}{\end{array}}
\newcommand{\bi}{\begin{itemize}}
\newcommand{\ei}{\end{itemize}}

\def\bea#1\eea{\allowdisplaybreaks \begin{align}#1\end{align}}
 \newcommand{\ben}{\begin{enumerate}}
\newcommand{\een}{\end{enumerate}}
\newcommand{\bean}{\begin{eqnarray*}}
\newcommand{\eean}{\end{eqnarray*}}
\newcommand{\eref}[1]{(\ref{#1})}

\newcommand{\BC}{\mathbb{C}}

\newcommand{\BZ}{\mathbb{Z}}

\definecolor{light-gray}{gray}{0.5}

\newcommand{\blue}{\color{blue}}

\newcommand{\red}{\color{red}}

\def\aup#1 {\overset{#1}{\uparrow} \, \overset{\tilde{#1}}{\downarrow}}

\tikzset{snake it/.style={decorate, decoration={snake, amplitude=.4mm, segment length=2mm,
                       post length=0mm,pre length=0mm}}}
                       
 \newcommand{\GCD}{\mathrm{GCD}}

\hypersetup{
	pdftitle={Comments on Non-invertible Symmetries in Argyres-Douglas Theories},    
	pdfauthor={\textcopyright\ Federico Carta, Simone Giacomelli, Noppadol Mekareeya, Alessandro Mininno},     
	pdfsubject={HEP},   
	pdfcreator={pdfLaTex},   
	pdfproducer={LaTex}, 
	pdfkeywords={},
	colorlinks=true,
}

\makeatletter
\newsavebox{\measure@tikzpicture}
\NewEnviron{scaletikzpicturetowidth}[1]{%
  \def\tikz@width{#1}%
  \begin{lrbox}{\measure@tikzpicture}%
  \BODY
  \end{lrbox}%
  \pgfmathparse{#1/\wd\measure@tikzpicture}%
  \BODY
}
\makeatother

\makeatletter
\def\squarecorner#1{
    \pgf@x=\the\wd\pgfnodeparttextbox%
    \pgfmathsetlength\pgf@xc{\pgfkeysvalueof{/pgf/inner xsep}}%
    \advance\pgf@x by 2\pgf@xc%
    \pgfmathsetlength\pgf@xb{\pgfkeysvalueof{/pgf/minimum width}}%
    \ifdim\pgf@x<\pgf@xb%
        \pgf@x=\pgf@xb%
    \fi%
    \pgf@y=\ht\pgfnodeparttextbox%
    \advance\pgf@y by\dp\pgfnodeparttextbox%
    \pgfmathsetlength\pgf@yc{\pgfkeysvalueof{/pgf/inner ysep}}%
    \advance\pgf@y by 2\pgf@yc%
    \pgfmathsetlength\pgf@yb{\pgfkeysvalueof{/pgf/minimum height}}%
    \ifdim\pgf@y<\pgf@yb%
        \pgf@y=\pgf@yb%
    \fi%
    \ifdim\pgf@x<\pgf@y%
        \pgf@x=\pgf@y%
    \else
        \pgf@y=\pgf@x%
    \fi
    \pgf@x=#1.5\pgf@x%
    \advance\pgf@x by.5\wd\pgfnodeparttextbox%
    \pgfmathsetlength\pgf@xa{\pgfkeysvalueof{/pgf/outer xsep}}%
    \advance\pgf@x by#1\pgf@xa%
    \pgf@y=#1.5\pgf@y%
    \advance\pgf@y by-.5\dp\pgfnodeparttextbox%
    \advance\pgf@y by.5\ht\pgfnodeparttextbox%
    \pgfmathsetlength\pgf@ya{\pgfkeysvalueof{/pgf/outer ysep}}%
    \advance\pgf@y by#1\pgf@ya%
}
\makeatother

\pgfdeclareshape{square}{
    \savedanchor\northeast{\squarecorner{}}
    \savedanchor\southwest{\squarecorner{-}}

    \foreach \x in {east,west} \foreach \y in {north,mid,base,south} {
        \inheritanchor[from=rectangle]{\y\space\x}
    }
    \foreach \x in {east,west,north,mid,base,south,center,text} {
        \inheritanchor[from=rectangle]{\x}
    }
    \inheritanchorborder[from=rectangle]
    \inheritbackgroundpath[from=rectangle]
}

\tikzset{stretch/.initial=1}
\newcommand\drawloop[4][]%
   {\draw[shorten <=0pt, shorten >=0pt,#1]
      ($(#2)!\pgfkeysvalueof{/tikz/stretch}!(#2.#3)$)
      let \p1=($(#2.center)!\pgfkeysvalueof{/tikz/stretch}!(#2.north)-(#2)$),
          \n1= {veclen(\x1,\y1)*sin(0.5*(#4-#3))/sin(0.5*(180-#4+#3))}
      in arc [start angle={#3-90}, end angle={#4+90}, radius=\n1]%
   }

\preprint{ZMP-HH/23-3}
\title{Comments on Non-invertible Symmetries in Argyres-Douglas Theories}
\author[a]{Federico Carta,}
\author[b]{~Simone Giacomelli,}
\author[c,d]{~Noppadol Mekareeya}
\author[e]{\\ and Alessandro Mininno}
\affiliation[a]{Department of Mathematical Sciences,
		Durham University, \\ Durham, DH1 3LE, United Kingdom}
\affiliation[b]{Dipartimento di Fisica, Universit\`a di Milano-Bicocca, Piazza della Scienza 3, \\ I-20126 Milano, Italy}
\affiliation[c]{INFN, sezione di Milano-Bicocca, Piazza della Scienza 3, \\ I-20126 Milano, Italy}
\affiliation[d]{Department of Physics, Faculty of Science, Chulalongkorn University, \\ Phayathai Road,
Pathumwan, Bangkok 10330, Thailand}
\affiliation[e]{II. Institut f\"ur Theoretische Physik, Universit\"at Hamburg,\\
Luruper Chaussee 149, 22607 Hamburg, Germany}
\emailAdd{federico.carta@durham.ac.uk}
\emailAdd{simone.giacomelli@unimib.it}
\emailAdd{n.mekareeya@gmail.com}
\emailAdd{alessandro.mininno@desy.de}
\abstract{
We demonstrate the presence of non-invertible symmetries in an infinite family of superconformal Argyres-Douglas theories. This class of theories arises from diagonal gauging of the flavor symmetry of a collection of multiple copies of $D_p(\mathrm{SU}(N))$ theories. The same set of theories that we study can also be realized from 6d $\mathcal{N}=(1,0)$ compactification on a torus. The main example in this class is the $(A_2, D_4)$ theory. We show in detail that this specific theory bears the same structures of non-invertible duality and triality defects as those of $\CN=4$ super Yang-Mills with gauge algebra $\mathfrak{su}(2)$. We extend this result to infinitely many other Argyres-Douglas theories in the same family, including those with central charges $a=c$ whose conformal manifold is one dimensional, and those with $a\neq c$ whose conformal manifold has dimension larger than one. Our result is supported by examining certain special cases that can be realized in terms of theories of class $\mathcal{S}$.}

\allowdisplaybreaks[1]

\begin{document} 

\maketitle

\section{Introduction}

In the past years, there has been much interest in the study of generalized global symmetries in quantum field theory (QFT) \cite{Gaiotto:2014kfa}. A general way to understand symmetries involves the study of topological operators, also called defects, which are supported on generic submanifolds $M$ of spacetime. These defects are associated with the symmetry itself and implement its action on the other operators of the QFT. For instance, a 0-form symmetry $G$ acts on point operators, and is described as generated by a set of codimension-1 topological operators $U_g(M)$ (one per group element $g\in G$). The action of the 0-form symmetry on a local point operator can be thought as encircling such operator with the topological one, and then shrinking the latter to zero size.
Higher-form symmetries, both continuous and discrete, admit an analogous description where the topological defects are supported on submanifolds of higher codimension \cite{Kapustin:2014gua,Gaiotto:2014kfa}.

Upon composition, topological defects can satisfy group-like laws (as in the case of the higher-form symmetries we just mentioned above) or more general fusion properties. For this reason, one can distinguish between invertible topological defects, and non-invertible ones. Thus one can distinguish between invertible and non-invertible symmetries.

There are many examples in the literature of non-invertible symmetries in the context of 2d or 3d QFTs, see for example \cite{Verlinde:1988sn,Petkova:2000ip,Fuchs:2002cm,Frohlich:2004ef,Bachas:2004sy,Frohlich:2006ch,Frohlich:2009gb,Bachas:2009mc,Kapustin:2010if,Carqueville:2012dk,Brunner:2013xna,Tachikawa:2017gyf,Bhardwaj:2017xup,Chang:2018iay,Lin:2019hks,Thorngren:2019iar,Gaiotto:2019xmp,Gaiotto:2020iye,Komargodski:2020mxz,Kaidi:2021gbs,Burbano:2021loy,Huang:2021nvb,Thorngren:2021yso,Huang:2021zvu}. More recently, they have also been studied in higher dimensional theories \cite{Wan:2018bns,Choi:2021kmx,Wang:2021vki,Apruzzi:2021nmk,Wang:2021ayd,Kaidi:2021xfk,Heidenreich:2021xpr,Koide:2021zxj,Roumpedakis:2022aik,Damia:2022bcd,Kaidi:2022cpf,Niro:2022ctq,Antinucci:2022eat,Hayashi:2022fkw,Benini:2022hzx,Cordova:2022ieu,Choi:2022jqy,Bashmakov:2022jtl,Bhardwaj:2022lsg,Bartsch:2022mpm,Heckman:2022muc,Arias-Tamargo:2022nlf,Freed:2022qnc,Apruzzi:2022rei,Cordova:2022rer,Choi:2022rfe,Damia:2022rxw,Kaidi:2022uux,Antinucci:2022vyk,GarciaEtxebarria:2022vzq,Lin:2022xod,Bhardwaj:2022yxj,Choi:2022zal,Kaidi:2023maf,Garcia-Valdecasas:2023mis, Antinucci:2022cdi,Lin:2022dhv,Choi:2022fgx,Cordova:2022fhg,Das:2022fho,GarciaEtxebarria:2022jky,Karasik:2022kkq,Bhardwaj:2022kot,Inamura:2022lun,Bhardwaj:2022maz,Mekareeya:2022spm,Bashmakov:2022uek,Lu:2022ver,Giaccari:2022xgs,Heckman:2022xgu,Apte:2022xtu,Bartsch:2022ytj,Yokokura:2022alv,Putrov:2023jqi},\footnote{See also recent reviews in \cite{Argyres:2022mnu,Cordova:2022ruw}.} including several supersymmetric models such as $\CN=4$ super Yang-Mills (SYM) and theories of class $\CS$ in 4d.

One remarkable class of supersymmetric QFTs, for which the possibility of existence of non-invertible symmetries has yet to be explored, consists of Argyres-Douglas (AD) theories. These are 4d $\mathcal{N}=2$ superconformal field theories (SCFTs) which enjoy many interesting features, such as fractional dimensions of the Coulomb branch (CB) operators. Theories of AD-type were historically first found in \cite{Argyres:1995jj,Argyres:1995xn,Eguchi:1996vu,Eguchi:1996ds} where they were seen to arise as the low-energy effective theory at very special points in the CB of an asymptotically free mother theory. At these specific points, non-local dyons become massless, making the AD theory intrinsically non-Lagrangian.\footnote{See also \cite{Gaiotto:2010jf,Giacomelli:2012ea,Akhond:2021xio} for reviews on 4d SCFTs.}  It was later understood that AD theories can be realized in many contexts, such as within the framework of class $\mathcal{S}$ \cite{Gaiotto:2009we}, or by type IIB geometric engineering on singular Calabi-Yau (CY) 3-fold \cite{Shapere:1999xr,Cecotti:2010fi,Xie:2015rpa}, or by twisted compactification of $6$d $\mathcal{N}=(1,0)$ theories \cite{DelZotto:2015rca,Ohmori:2015pua,Ohmori:2015pia}. In the present work we will be primarily concerned with the SCFTs of $D_p(G)$ type \cite{Cecotti:2012jx, Cecotti:2013lda, Giacomelli:2017ckh} and their gauging.  In the context of geometric engineering and class $\mathcal{S}$, it has been discovered that several of such AD theories possess non-trivial 1-form \cite{DelZotto:2020esg, Closset:2020scj, Hosseini:2021ged, Bhardwaj:2021mzl} and 2-group symmetries \cite{Bhardwaj:2021mzl, Carta:2022fxc}.  The presence of these symmetries brings about rich dynamical consequences that have been studied in our previous work \cite{Carta:2022spy, Carta:2022fxc}. 

In this article, we continue the exploration of generalized global symmetries in AD theories.  In particular, we focus on the presence of non-invertible symmetries in an infinite family of AD theories arising from compactification of 6d $\CN=(1,0)$ theories on a torus, which admit a description in terms of diagonal gauging of the flavor symmetry of a certain number of $D_p(\SU(N))$ AD theories.  One of the notable non-Lagrangian models in this class is the $(A_2, D_4)$ theory, which can be realized from diagonal gauging of the $\su(2)$ flavor symmetry algebra of  three copies of the $D_3(\SU(2)) = (A_1,A_3)$ theory \cite{Buican:2016arp, Closset:2020scj, Closset:2020afy} and possesses the central charges that satisfy $a=c$. We point out that this model admits non-invertible duality and triality defects as for the 4d $\CN=4$ super Yang-Mills (SYM) theories \cite{Kaidi:2021xfk, Choi:2021kmx, Choi:2022zal, Kaidi:2022uux} with gauge algebra $\su(2)$, upon choosing an appropriate global form of $\su(2)$.  In this paper, we point out that other AD theories with central charges $a=c$ in this class can be discussed in parallel to the $\CN=4$ SYM theories. Moreover, we extend this result to theories with $a \neq c$ by restricting ourselves to an appropriate one dimensional submanifold of the conformal manifold parametrized by the complex structure $\tau$ of the torus associated with compactification. The $\SL(2,\BZ)$ action inherited from the mapping class of the torus acts on $\tau$ in the usual way, and so there exist values of $\tau$ such that this is invariant under action of a certain element of $\SL(2,\BZ)$ which becomes a zero-form symmetry of the theory.  We observe that these theories once again reproduce the structure of the $\CN=4$ SYM. Our observation is supported by examining some special cases that can be realized as theories of class $\CS$.

\subsubsection*{Organization of the Paper}

This paper is organized as follows. In Section \ref{sec:6dN10theoriesonT2}, we consider 4d SCFTs obtained by the conformal gauging of the flavor symmetry algebra $\su(N)$ of a collection of $D_p(\SU(N))$ theories. We review that these theories can also be obtained from a $T^2$ compactification of 6d $\mathcal{N}=(1,0)$ theories realized by putting F-theory on the orbifold $(\BC^2\times T^2)/\Gamma$, where $\Gamma$ is a $\ZZ_k$ cyclic subgroup of $\SU(3)$. We explain details of the orbifold and analyze the higher-form symmetries for the 6d theories in Section \ref{sec:higherformsymmetries}. In Section \ref{sec:triniontheories} we discuss the descriptions of the 4d SCFTs in detail. The 1-form symmetry of the 4d theories is derived in this section, utilizing information in Appendix \ref{sec:1formtrinion}, where the operators of each $D_p(G)$ theory involved in the conformal gauging are analyzed by means of the Lagrangian description of the theory dimensionally reduced to 3d.  

The main result of the paper is presented with an example in Section \ref{sec:A2D4}, where we show that, upon choosing an appropriate global form for the gauge group, the $(A_2,D_4)$ theory contains the non-invertible duality and triality defects.  We then generalize this result to infinite families of AD theories in Section \ref{sec:generalization}, distinguishing between those that have equal central charges ($a=c$), and those which do not. If $a=c$, the non-invertible symmetries can be established in an analogous way to that of the 4d $\mathcal{N}=4$ SYM \cite{Kaidi:2021xfk, Choi:2021kmx, Choi:2022zal, Kaidi:2022uux}. On the other hand, if $a\neq c$, a similar result is established provided that one restricts oneself to an appropriate one dimensional submanifold of the conformal manifold as mentioned above. This procedure is demonstrated in Section \ref{sec:Sdualitytrionion}, where we present certain examples that admit a class $\CS$ description.

\section{6d \texorpdfstring{$\mathcal{N}=(1,0)$}{} Theories on \texorpdfstring{$T^2$}{}} 
\label{sec:6dN10theoriesonT2}

In this work, we are mainly interested in ``trinion'' theories obtained by conformally gauging the flavor symmetry algebra $\fg$ associated to the group $G$ of a collection of $D_p(G)$ theories. For completeness, let us briefly summarize the information about the $D_p(G)$ theory as follows.

The $D_p(G)$ theories\footnote{They are also known as $D_p^b(G)$ with $b = h(G)$, i.e. the Coxeter number of $G$, where in this case the superscript $b$ is conventionally omitted.} can be realized in class $\mathcal{S}$ \cite{Wang:2018gvb,Gaiotto:2009we,Gaiotto:2009hg,Xie:2012hs} as a compactification of a 6d $\mathcal{N}=(2,0)$ theory on a sphere with a regular full puncture and an irregular puncture. Whenever $G$ is not simply-laced, one needs to consider outer-automorphism twists in this construction. The theories also admit another realization in terms of Type IIB geometric engineering \cite{Shapere:1999xr,Cecotti:2010fi} on a non-compact CY 3-fold realized as the zero-locus of a single hypersurface singularity in $\mathbb{C}^3\times \mathbb{C}^*$. For example, the hypersurface singularity realizing the $D_p(\SU(N))$ theory, which is the protagonist of this paper, in Type IIB is given by
\begin{equation}
    W(x_1,x_2,x_3,z) = x_1^2 + x_2^N + x_3^2 + z^{p+1} = 0\coma \text{with } \quad \Omega = \frac{dx_1 dx_2 dx_3 dz}{zdW}\coma
\end{equation}
where $z$ is the $\mathbb{C}^*$ variable. By closure of the regular puncture in the class $\CS$ realization of the $D_p(G)$ theory, the theory has a realization in Type IIB on non-compact CY 3-fold given by the zero-locus of a single hypersurface singularity, this time, in $\mathbb{C}^4$. Conventionally, when $G =A_n$, $D_n$ or $E_n$, the resulting theory can be identified with the $(A_{p-h(G)-1},G)$ AD theory \cite{Cecotti:2010fi}.

As is well known, the vanishing of the $G$ beta function imposes a constraint on the set of $D_p(G)$ models we can choose: We are allowed to consider either 4 copies of $D_2(G)$ or three theories with $(p_1,p_2,p_3)=(3,\,3,\,3)$, $(p_1,p_2,p_3)=(4,\,4,\,2)$ or $(p_1,p_2,p_3)=(2,\,3,\,6)$ \cite{Cecotti:2013lda}. In \cite{DelZotto:2015rca}, it was pointed out that such models can be realized by compactifying certain 6d $\mathcal{N}=(1,0)$ theories on $T^2$. More precisely, the double dimensional reduction of the 6d theory leads to a (generically) non-conformal 4d model (which we refer to as KK theory from now on) and by tuning Coulomb branch and mass parameters in this theory we can find singular points in the moduli space hosting the $\mathcal{N}=2$ SCFTs we are after. This procedure is analogous to the technique originally used to construct the first examples of Argyres-Douglas theories, which are realized at singular points in the Coulomb branch of asymptotically free gauge theories \cite{Argyres:1995jj}. We will now review the argument of \cite{DelZotto:2015rca} and describe the relevant 6d theories.

One convenient way for defining a 6d SCFT is to specify the Calabi-Yau 3-fold, which engineers it in F-theory. In order to get the trinion theories we are after, we can focus on Calabi-Yau manifolds which are orbifolds of $\mathbb{C}^2\times T^2$. The orbifold group $\Gamma$ is a discrete subgroup of $\SU(3)$ and corresponds to the direct product of a cyclic group $\mathbb{Z}_k$ acting as 
\begin{equation} 
(\omega_k,\omega_k;\omega_k^{-2})\coma
\end{equation} 
where $\omega_k^{-2}$ acts on $T^2$ (this indeed restricts the possible values of $k$ to the set $k=2,\,3,\,4,\,6,\,8,\,12$) and a ADE discrete subgroup of $\SU(2)$ $\Gamma_{ADE}$ acting on $\mathbb{C}^2$ only. It will turn out that the $G$ gauge algebra of the trinion theory is determined by the choice of ADE group appearing in the orbifold. The cyclic group $\mathbb{Z}_k$ in turn determines the value of the $p_i$ parameters, therefore specifying the actual $D_p(G)$ theories involved in the gauging, as follows
\begin{equation}
\begin{array}{c|c}
\text{Value of $k$} & \text{value of $p_i$ parameters}\\
\hhline{=|=}
3,\,6 & p_1=p_2=p_3=3 \\
\hline 
4 & p_1=p_2=p_3=p_4=2 \\
\hline 
8 & p_1=p_2=4;\; p_3=2 \\
\hline 
12 & p_1=2;\; p_2=3;\; p_3=6 
\end{array}
\label{tablep}
\end{equation}
We neglected the case $k=2$ which correspond to a trivial action on the torus and therefore reduces to a ADE orbifold of $\mathbb{C}^2$. This of course leads to the $\mathcal{N}=(2,0)$ theories in 6d.

Let us focus on the special case in which $\Gamma_{ADE}$ is also cyclic of order $n$, in which case the group $G$ will be unitary. The group action on $\mathbb{C}^2\times T^2$ is of course
$$(\omega_n,\omega_n^{-1};1)\coma$$ 
where the action on the $T^2$ is trivial. 
Overall, the orbifold group acts as follows: 
\begin{equation}\label{orbidef} \mathbb{Z}_k: \;(\omega_k,\omega_k;\omega_k^{-2})\,; \quad \mathbb{Z}_n: \; (\omega_n,\omega_n^{-1};1)\fstop \end{equation}
We will now analyze the two extreme cases in which $k$ and $n$ are coprime or $k$ is a divisor of $n$. 

\subsubsection*{\texorpdfstring{$k$}{} and \texorpdfstring{$n$}{} Coprime} 

In this case, the orbifold group is cyclic of order $nk$ and the 6d theory turns out to be a non-Higgsable theory, which is entirely specified by the configuration of curves in the $\mathbb{C}^2$ base. Exploiting the results of \cite{Heckman:2013pva}, we conclude that the curves in the base form a linear chain and their self-intersection $\{n_i\}_{i=1,\dots,r}$ is encoded in the continuous fraction 
\begin{equation}\label{contfrac}
\frac{nk}{q}=n_1-\frac{1}{n_2-\frac{1}{n_3-\dots}}\coma
\end{equation}
where $q$ satisfies the relation 
\begin{equation}\label{qequation}
q(n+k)=(n-k)\,\mod \, nk\fstop
\end{equation} 
The latter can always be solved, and the corresponding curve self-intersections are reported in \cite{DelZotto:2015rca}. 
In the special case $n=1$, \eqref{qequation} is solved by $q=1$ and the resulting 6d theory is a minimal model with a single curve of self-intersection $-k$.

The origin of \eqref{qequation} is as follows: In \cite{Heckman:2013pva}, it was shown that for all non-Higgsable theories the underlying Calabi-Yau 3-fold is always an orbifold of $\mathbb{C}^2\times T^2$ and in the case of a linear chain of curves in the base the orbifold action is
\be\label{orbi2}
(\omega_p,\omega_p^q; \omega_p^{-q-1})\coma
\ee 
with $p$ and $q$ coprime.
The pattern of self-intersections is encoded in a continuous fraction analogous to \eqref{contfrac} with $p/q$ on the left-hand side. We should therefore first try to rewrite \eqref{orbidef} in the form \eqref{orbi2}. Since for $n$ and $k$ coprime \eqref{orbidef} is equivalent to a cyclic group of order $nk$, we immediately get $p=nk$. We should then demand that $(\omega_k\omega_n,\omega_k\omega_n^{-1})$ is equivalent to $(\omega_p,\omega_p^q)$, which leads to \eqref{qequation}. 

\subsubsection*{\texorpdfstring{$n$}{} Multiple of \texorpdfstring{$k$}{}} 

In this case, it turns out that the 6d theory engineered in F-theory is a non-minimal conformal matter \cite{DelZotto:2014hpa}, which can be alternatively constructed by placing multiple M5-branes at an orbifold singularity in M-theory. We will denote this theory as $\mathcal{T}(G,N)$ where $G$ denotes the orbifold type and $N$ is the number of M5-branes. In order to understand this statement, let us go back to \eqref{orbidef} and denote the generators of $\mathbb{Z}_k$ and $\mathbb{Z}_n$ by $g$ and $h$ respectively for convenience. We can notice that $gh^{n/k}$ and $gh^{-n/k}$ act on $\mathbb{C}^2\times T^2$ in the following way: 
$$gh^{n/k}: \;(\omega_k^2,1;\omega_k^{-2})\,; \quad gh^{-n/k}: \; (1,\omega_k^{2};\omega_k^{-2})\fstop$$
Here we recognize the singularities in F-theory produced by 7-branes of type $D_4$ for $k=4$, $E_6$ for $k=3,\,6$, $E_7$ for $k=8$ and $E_8$ for $k=12$. The groups $G$ associated to the 7-branes we have just listed become global symmetries of the 6d SCFTs, and therefore we expect to get models with $G\times G$ global symmetry. These correspond to the $(D_4,D_4)$, $(E_6,E_6)$, $(E_7,E_7)$ and $(E_8,E_8)$ conformal matters, respectively. The number of M5-branes can be determined by studying the singularity at which the two $G$-type non-compact divisors collide. It turns out to be $n/k$ for $k=3$ and $2n/k$ in the other cases. 

Another equivalent argument is as follows: The theories $\mathcal{T}(G,N)$ we have just discussed can be alternatively engineered with the orbifold 
\begin{equation}\label{orbidef2} 
\mathbb{Z}_r: \;(1,\omega_r^{-1};\omega_r)\,; \quad \mathbb{Z}_{Nr}: \; (\omega_{Nr},\omega_{Nr}^{-1};1)\fstop 
\end{equation} 
We find $G=D_4$ for $r=2$ and $G=E_6, E_7, E_8$ for $r=3,4,6$ respectively. We can now observe that for $k=3$ \eqref{orbidef} is equivalent to \eqref{orbidef2} with $r=3$ and $N=n/3$. For $k=4,6,8$ and $12$ instead, \eqref{orbidef} is equivalent to \eqref{orbidef2} with $r=k/2$ and $N=2n/k$.

\subsection{Higher-form Symmetries in 6d} 
\label{sec:higherformsymmetries}

Once the configuration of curves in the F-theory base is known, it is possible to identify the defect group of the 6d theory using the method proposed in \cite{DelZotto:2015isa}. Since higher-form symmetries do not change upon higgsing, we can reduce to considering the non-Higgsable theory our SCFT flows to. 

In the case of non-minimal conformal matter we have discussed before, this is the $\mathcal{N}=(2,0)$ theory living on the stack of M5-branes. In the case, the higgsing is realized geometrically in M-theory by moving the M5-branes away from the ADE orbifold singularity and therefore the 2-form symmetry is $\mathbb{Z}_{n/k}$ for $k=3$ and $\mathbb{Z}_{2n/k}$ in the other cases. 

The case of non-Higgsable theories we have seen before (when the order of the two orbifold groups $n$ and $k$ are coprime) can be determined as follows: Since the collection of curves in the base are arranged as a linear chain, with neighboring curves having intersection 1, according to \cite{DelZotto:2015isa} the 2-form symmetry is a cyclic group of order equal to the determinant of the adjacency matrix, whose entry $ij$ is minus the intersection of the $i$-th and $j$-th curves. This is always equal to the numerator appearing on the LHS of \eqref{contfrac} and therefore we conclude that the 2-form symmetry is $\mathbb{Z}_{nk}$. 

Finally, some of these theories also have a 1-form symmetry. This can be determined by moving on the tensor branch, where the theory becomes Lagrangian. The 1-form symmetry is therefore given by the center of the gauge group modulo screening due to the matter sector (both the hypermultiplets in the various representations of the gauge group and the BPS strings \cite{Apruzzi:2020zot, Bhardwaj:2020phs}). For example, in the case of minimal models ($n=1$ and $k=2,\,3,\,4,\,6,\,8,\,12$) we have a simple gauge group without matter fields and therefore the 1-form symmetry is simply the center of the gauge group: trivial for $k=2,\,12$, $\mathbb{Z}_3$ for $k=3,\,6$, $\mathbb{Z}_2\times \mathbb{Z}_2$ for $k=4$ and $\mathbb{Z}_2$ for $k=8$. The 1-form symmetry will not play a relevant role in the present work.

\subsection{Trinion Theories from 6d}
\label{sec:triniontheories}

Let us now discuss in more detail the 4d $\mathcal{N}=2$ SCFTs we get upon compactification on $T^2$ of the 6d theories we have described above. As was pointed out in \cite{Ohmori:2015pua, Ohmori:2015pia}, dimensional reduction of a 6d SCFT to 4d in general leads to a KK theory which is not conformal: it has a dynamically generated scale which coincides with the inverse size of the $T^2$. For example, the 4d reduction of non-minimal conformal matter analyzed in \cite{ Ohmori:2015pia} involves strongly-interacting sectors coupled through an infrared free vector multiplet. The only known cases in which upon dimensional reduction we get a superconformal theory is when the 6d theory has enhanced $\mathcal{N}=(2,0)$ supersymmetry, or it is very-Higgsable \cite{Ohmori:2015pua} (meaning it can be higgsed to a collection of free hypermultiplets). 

Since in this work we are considering SCFTs which are not very-Higgsable, we should not expect to get a SCFT directly upon double dimensional reduction but rather a non-conformal KK theory. The most natural question then is how we can describe it. The strategy adopted in \cite{DelZotto:2015rca} is to exploit string dualities: As we have seen 6d theories can be defined as compactification of F-theory on singular 3-folds, orbifolds of $\mathbb{C}^2\times T^2$ in our case, and therefore we can define the KK theory as Type IIB string theory\footnote{Putting the 6d theory on $T^2$ is equivalent to study F-theory on CY$ \times T^2 = \text{CY}\times S^1\times S^1$. This by definition is equivalent to M-theory on CY$\times S^1$ or Type IIA on CY. We then exploit mirror symmetry to move to a Type IIB description.} compactified on the mirror Calabi-Yau. This admits a Landau-Ginzburg (LG) type description (we will discuss this point more in detail later on) which allows us to understand the structure of the Coulomb branch of the KK theory.\footnote{We are grateful to Michele Del Zotto for the discussion on this point.} Once we have the Type IIB geometry available, we can try to identify singular points on the Coulomb branch hosting a SCFT in the infrared. This is analogous to the method first proposed in \cite{Argyres:1995jj, Argyres:1995xn} to identify singular points on the Coulomb branch of $\mathcal{N}=2$ gauge theories and study the SCFTs living at those points. 

In \cite{DelZotto:2015rca} it was argued that by tuning Coulomb branch moduli in the KK theory we can obtain the trinion SCFTs we are interested in. More precisely, starting from the orbifold Calabi-Yau manifolds \eqref{orbidef}, we can end up with a $\fg=\su$ vector multiplet coupled to several $D_p(G)$  theories, according to the pattern reported in \eqref{tablep}. All these theories are conformal and exhibit an exactly marginal coupling inherited from the complex structure of the torus we compactify the 6d theory on. When {\it $n$ is a multiple of $k$}, we find $\fg=\su(n)$ and can see from \eqref{tablep} that for all the $D_p(\SU(n))$ theories involved the parameter $p$ is a divisor of $n$. Theories with this property are known to be Lagrangian and describe a linear quiver with $p-1$ special unitary gauge groups \cite{Cecotti:2013lda}. Overall, we end up with a Lagrangian SCFT as listed below, where, for convenience of the presentation, we choose the global form associated with each $\su(n)$ gauge algebra to be $\SU(n)$:
    \begin{align}
       {{k=3}}:  & \text{\hspace{2.5cm}\begin{tikzpicture}[scale=0.86,font=\footnotesize,baseline=0.75cm]
\node (a1) at (0,0)    {$\SU(n)$};
\node (a2) at (0,2)    {$\SU(n/3)$};
\node (a3) at (0,1)    {$\SU(2n/3)$};
\node (a4) at (-4.2,0) {$\SU(n/3)$};
\node (a5) at (-2,0)   {$\SU(2n/3)$};
\node (a7) at (4.2,0)  {$\SU(n/3)$};
\node (a6) at (2,0)    {$\SU(2n/3)$};
\draw (a2)--(a3)--(a1) (a4)--(a5)--(a1)--(a6)--(a7);
\end{tikzpicture}}\nonumber \\
        {{k=4}}:   & \text{\hspace{5cm} \begin{tikzpicture}[scale=0.86,font=\footnotesize,baseline=0cm]
\node (A1) at (0,0) {$\SU(n)$};
\node (A2) at (1.2,-1.2) {$\SU(n/2)$};
\node (A3) at (-1.2,1.2) {$\SU(n/2)$};
\node (A4) at (1.2,1.2) {$\SU(n/2)$};
\node (A5) at (-1.2,-1.2) {$\SU(n/2)$}; 
\draw (A2)--(A1) (A3)--(A1) (A4)--(A1) (A5)--(A1); 
\end{tikzpicture} } \nonumber \\
        {{k=6}}:   & \text{\hspace{2.5cm} \begin{tikzpicture}[scale=0.86,font=\footnotesize,baseline=0.75cm] 
\node (B1) at (0,0)    {$\SU(n)$};
\node (B2) at (0,2)    {$\SU(n/3)$};
\node (B3) at (0,1)    {$\SU(2n/3)$};
\node (B4) at (-4.2,0) {$\SU(n/3)$};
\node (B5) at (-2,0)   {$\SU(2n/3)$};
\node (B7) at (4.2,0)  {$\SU(n/3)$};
\node (B6) at (2,0)    {$\SU(2n/3)$};
\draw (B2)--(B3)--(B1) (B4)--(B5)--(B1)--(B6)--(B7);
\end{tikzpicture} } \label{eq:LagrangianSCFTs}\\
        {{k=8}}:   & \text{\hspace{0.6cm} \begin{tikzpicture}[scale=0.86,font=\footnotesize,baseline=0.5cm]
\node (C1) at (0,0)     {$\SU(n)$};
\node (C2) at (0,1)     {$\SU(n/2)$};
\node (C3) at (-6.4,0)  {$\SU(n/4)$};
\node (C4) at (-4.2,0)  {$\SU(n/2)$};
\node (C5) at (-2,0)    {$\SU(3n/4)$};
\node (C7) at (4.2,0)   {$\SU(n/2)$};
\node (C6) at (2,0)     {$\SU(3n/4)$};
\node (C8) at (6.4,0)   {$\SU(n/4)$};
\draw (C2)--(C1) (C3)--(C4)--(C5)--(C1)--(C6)--(C7)--(C8);
\end{tikzpicture} }\nonumber \\
        {{k=12}}:  & \text{ \begin{tikzpicture}[scale=0.86,font=\footnotesize,baseline=0.75cm]
\node (D1) at (5,0)  {$\SU(n)$};
\node (D2) at (5,2)  {$\SU(n/3)$};
\node (D3) at (5,1)    {$\SU(2n/3)$};
\node (D4) at (-6.4,0) {$\SU(n/6)$};
\node (D5) at (-4.2,0) {$\SU(n/3)$};
\node (D8) at (-2,0)   {$\SU(n/2)$};
\node (D6) at (0.4,0)  {$\SU(2n/3)$};
\node (D7) at (2.8,0)    {$\SU(5n/6)$};
\node (D9) at (7.2,0)  {$\SU(n/2)$};
\draw (D2)--(D3)--(D1) (D4)--(D5)--(D8)--(D6)--(D7)--(D1)--(D9);
\end{tikzpicture} }\nonumber
    \end{align}
Notice that for all the quivers, the 1-form symmetry of the 4d theory (namely $\BZ_{n/3}$ for $k=3$ and $\BZ_{2n/k}$ for other cases) is equal to the 2-form symmetry of the parent 6d theory.  We may gauge such a 1-form symmetry or its subgroup in each of the above theories.

In the opposite case when $n$ and $k$ in \eqref{orbidef} are coprime the situation is a bit more involved. The vector multiplet of the trinion theory corresponds to $\su(n)$ for $k=3$ and to $\su(2n)$ in the other cases. We can easily see from \eqref{tablep} that for $k=3,\,6$ (when all the $D_p(\SU)$ theories involved have $p=3$) $p$ is coprime with $n$. In this case it is known that the global symmetry of the $D_p(G)$ theory is exactly $G$, without further enhancements, and that the theory is isolated, meaning it has no marginal couplings. As a result, the conformal manifold of the trinion theory is one-dimensional and is inherited from the 6d torus, as we have mentioned before. For $k=4,\,8,\,12$ instead the trinion theory involves $D_p(\SU(2n))$ sectors with $p$ even and these all have (see \cite{Giacomelli:2020ryy}) one exactly marginal parameter and an extra $\U(1)$ global symmetry (besides the $\SU(2n)$ flavor symmetry). Overall, the dimension of the conformal manifold is $5$, $4$ and $3$ for $k=4$, $k=8$ and $k=12$ respectively. We find the following trinion\footnote{\label{footnote:quadrinion}For $k=4$, the theory actually arises from gauging of four $D_p(G)$ theories. It would be more appropriate to call it ``quadrinion". Nevertheless, for simplicity's sake, we abuse the terminology by referring to all of these quivers as trinions.} models for {\it $n$ and $k$ coprime}:

   \begin{align} 
       {{k=3}}:  & \text{\hspace{2.5cm}\begin{tikzpicture}[scale=0.86,font=\footnotesize,baseline=0.25cm]
\node (a1) at (0,0)    {$\SU(n)$};
\node (a3) at (0,1)    {$D_3(\SU(n))$};
\node (a4) at (-2.2,0) {$D_3(\SU(n))$};
\node (a6) at (2.2,0)  {$D_3(\SU(n))$};
\draw (a3)--(a1) (a4)--(a1) (a6)--(a1);
\end{tikzpicture}}\nonumber \\
        {{k=4}}:   & \text{\hspace{3.25cm} \begin{tikzpicture}[scale=0.86,font=\footnotesize,baseline=0cm]
\node (A1) at (0,0)  {$\SU(2n)$};
\node (A2) at (1.2,1.2) {$D_2(\SU(2n))$};
\node (A3) at (1.2,-1.2) {$D_2(\SU(2n))$};
\node (A4) at (-1.2,1.2)  {$D_2(\SU(2n))$};
\node (A5) at (-1.2,-1.2)  {$D_2(\SU(2n))$};
\draw (A2)--(A1) (A3)--(A1) (A4)--(A1) (A5)--(A1); 
\end{tikzpicture} } \nonumber \\
        {{k=6}}:   & \text{\hspace{2.25cm} \begin{tikzpicture}[scale=0.86,font=\footnotesize,baseline=0.25cm]
\node (B1) at (0,0)    {$\SU(2n)$};
\node (B3) at (0,1)    {$D_3(\SU(2n))$};
\node (B5) at (-2.3,0) {$D_3(\SU(2n))$};
\node (B6) at (2.3,0)  {$D_3(\SU(2n))$};
\draw (B3)--(B1) (B5)--(B1) (B6)--(B1);
\end{tikzpicture} } \label{eq:DpgaugingSCFTs}\\
        {{k=8}}:   & \text{\hspace{2.25cm} \begin{tikzpicture}[scale=0.86,font=\footnotesize,baseline=0.25cm]
\node (C1) at (0,0)    {$\SU(2n)$};
\node (C2) at (0,1)    {$D_2(\SU(2n))$};
\node (C5) at (-2.3,0) {$D_4(\SU(2n))$};
\node (C6) at (2.3,0)  {$D_4(\SU(2n))$};
\draw (C2)--(C1) (C1)--(C5) (C1)--(C6);
\end{tikzpicture} }\nonumber \\
        {{k=12}}:  & \text{\hspace{2.25cm} \begin{tikzpicture}[scale=0.86,font=\footnotesize,baseline=0.25cm]
\node (D1) at (0,0)    {$\SU(2n)$};
\node (D2) at (0,1)    {$D_3(\SU(2n))$};
\node (D3) at (-2.3,0) {$D_2(\SU(2n))$};
\node (D4) at (2.3,0)  {$D_6(\SU(2n))$};
\draw (D2)--(D1) (D4)--(D1) (D3)--(D1);
\end{tikzpicture} }\nonumber
    \end{align}
where, for the convenience of presentation, we have taken the global form of the $\su(n)$ and $\su(2n)$ gauge algebras to be respectively $\SU(n)$ and $\SU(2n)$.  Of course, other global forms can be chosen. We will consider the other choices in subsequent sections.

Since the condition that $n$ and $k$ are coprime implies that all the $p_i$ parameters as well are coprime with $n$, we find that for $k>2$, the 1-form symmetry is always $\mathbb{Z}_n$ apart from the case $k=6$, when the 1-form symmetry becomes $\mathbb{Z}_{2n}$.  The former is due to the screening effect of certain operators in the $D_{p_i}(\SU(2n))$ theories; we discuss this in detail in Appendix \ref{sec:1formtrinion}. Notice that this is a proper subgroup of the 2-form symmetry of the parent 6d theory, leading to the conclusion that the KK theory has a $\mathbb{Z}_k$ ($\mathbb{Z}_3$ for $k=6$) 1-form symmetry which does not act on the SCFT.   We also emphasize that in the cases in which all the $p_i$ parameters are coprime to the number of colors $N$ of the central node $\SU(N)$, the theory has central charges $a=c$.  To put it another way, whenever the $\BZ_{2n}$ center of the $\SU(2n)$ gauge group is broken to $\BZ_n$ due to the screening effect, the theory has central charges $a \neq c$.

Let us discuss more in detail the SCFTs one gets in this way starting from minimal models in 6d with $n=1$ and $k=2,\,3,\,4,\,6,\,8,\,12$: For $k=2$ the orbifold does not act at all on the $T^2$ and therefore we find in 6d the $\mathcal{N}=(2,0)$ theory of type $A_1$, whose $T^2$ reduction gives $\mathcal{N}=4$ SYM with gauge group $\SU(2)$. For $k=3$ the theory cannot be interpreted as a trinion since the gauge group at the center would be $\SU(1)$. In this case, it turns out that the SCFT can be interpreted as $D_3(\SU(2))$ \cite{DelZotto:2015rca}. In all other cases, we have $\SU(2)$ trinions. Since $D_2(\SU(2))$ is equivalent to a doublet of $\SU(2)$, the 1-form symmetry is trivial whenever we have a $D_2(\SU(2))$ sector and is $\mathbb{Z}_2$ otherwise. The cases with trivial 1-form symmetry, namely $k=4,\,8,\,12$ correspond to $\SU(2)$ SQCD with four flavors, $(A_3,A_3)$ and $(A_2,A_5)$ AD theories respectively \cite{Cecotti:2010fi}. The only case with $\mathbb{Z}_2$ 1-form symmetry is $k=6$, which corresponds to the $(A_2,D_4)$ AD theory \cite{Closset:2020afy}. This model will be the focus of Section \ref{sec:A2D4}. 

\subsection{Mirror Calabi-Yau and Trinion Theories}

Let us now explain how to derive \eqref{eq:DpgaugingSCFTs}. The strategy of \cite{DelZotto:2015rca} involves analyzing the Coulomb branch geometry of the KK theory $\CT_{\text{KK}}$ via its Calabi-Yau geometry (or Landau-Ginzburg superpotential) which is the mirror dual of the $T^2 \times \BC^2$ orbifold we have discussed so far. We would now like to explain how the mirror dual is constructed, focusing on the case in which $n$ and $k$ are coprime. 

It is easier to use the orbifold description \eqref{orbi2} where $p=nk$ and $q$ satisfies \eqref{qequation}. This group includes a $\BZ_r$ action on $T^2$, where $r=k$ for $k=3$ and $r=k/2$ for $k$ even. It is known that the mirror of a $\BZ_r$ orbifold of $T^2$ is given by the Landau-Ginzburg superpotential 
\be\label{t2orbi}\begin{array}{c|c}
\text{Orbifold order $r$} & \text{LG superpotential $W_{T^2/\BZ_r}(x_1, x_2, x_3)$} \\
\hhline{=|=}
3 & x_1^3+x_2^3+x_3^3+\tau x_1x_2x_3\\
\hline 
4 & x_1^2+x_2^4+x_3^4+\tau x_1x_2x_3\\ 
\hline 
6 & x_1^2+x_2^3+x_3^6+\tau x_1x_2x_3 
\end{array}
\ee 
where the $x_i$'s are complex variables and the parameter $\tau$ (the complex structure of the mirror $T^2$) will eventually be interpreted as the complex structure of the torus we compactify the 6d SCFT on. Apart from the $\tau$ term these LG models have in their chiral ring operators of dimension $\ell/r$ with $0\leq \ell<r$, associated with deformations of the geometry \eqref{t2orbi}, and their multiplicity is as follows: 
\be\label{specLG}\begin{array}{c|c|c|c|c|c|c}
\text{Orbifold order $r$} & \ell=0 & \ell=1 & \ell=2 & \ell=3 & \ell=4 & \ell=5 \\
\hhline{=|=|=|=|=|=|=}
2 & 1 & 4 &  &  &  & \\
\hline 
3 & 1 & 3 & 3 &  &  & \\
\hline 
4 & 1 & 2 & 3 & 2 &  &\\ 
\hline 
6 & 1 & 1 & 2 & 2 & 2 & 1
\end{array}
\ee 
From now on, we will denote the corresponding deformations as $\varphi_{a,\ell/r}(x_i)$. Note that in \eqref{specLG} we have included the spectrum for the $\BZ_2$ orbifold, even though we have not included the corresponding superpotential in \eqref{t2orbi} since there is no simple expression for it. We would like to stress that the analysis below will apply to the $r=2$ case as well, using the multiplicities reported in \eqref{specLG}.

The orbifold action on $\BC^2$ is accounted for by introducing two $\BC^*$ variables $y_1$ and $y_2$ and adding to the LG superpotential monomials of the form 
$$y_1^p \coma  y_2^p \coma  y_1^{[mr_1]_p}y_2^{[mr_2]_p}
\coma $$ 
where $r_1$ and $r_2$ denote the charges of the two complex coordinates under the orbifold group (in the case of interest for us $1$ and $q$ respectively), $[\cdot]_p$ denotes the $(\mod\, p)$ operation and $m$ is an integer $0\leq m<p$ subject to the constraint that all the terms appearing in the superpotential should be homogeneous of dimension one. Putting everything together we find the expression 
\be \label{WTK1}
W_{\CT_{\text{KK}}} = W_{T^2/\BZ_r} + y_1^p+ y_2^p + \sum_a\sum_{\ell}\sum_m \beta_{a,\ell,m}\varphi_{a,\ell/r}(x_i)y_1^m y_2^{[mq]_p}\fstop
\ee
The homogeneity constraint explicitly reads 
\be\label{mconstr} 
\frac{m+[mq]_p}{p}+\frac{\ell}{r}=1\fstop
\ee 
The parameters $\beta_{a,\ell,m}$ in \eqref{WTK1} describe the Coulomb branch moduli of the KK theory and are inherited from the 6d vector and tensor multiplets. We, therefore, conclude that \eqref{WTK1} encodes the structure of the Coulomb branch of our KK theory $\CT_{\text{KK}}$. 

If we now want to identify the trinion SCFTs we are after, it is convenient to work in terms of the mirror Calabi-Yau geometry in type IIB. This is given by the equation $$W_{\CT_{\text{KK}}} = 0$$ 
defined in the projectivization of the $(x_i, y_i)$-space. We can then go to the patch $y_2=1$ and introduce a new $\BC$ variable $w=y_1^b+1$ (where $b$ is an integer) in order to get a hypersurface in $\BC^4$ $W_{\CT_{\text{KK}}}(x_{1,2,3},w) = 0$, which is the usual description of the trinion SCFTs we are interested in. 
It is now convenient to notice that \eqref{mconstr} requires $m$ being a multiple of $r$ if we set $\ell=0$. This can be seen as follows: Eq. \eqref{qequation} implies that $q-1$ is a multiple of $k$ and $q+1$ is a multiple of $n$. The last statement implies that \eqref{mconstr} is satisfied when $m$ is a multiple of $k$ for $\ell=0$. On the other hand, when $k$ is even both $n$ and $q$ are necessarily odd, since they are both coprime with $k$. This implies that $q+1$ is an even multiple of $n$ and, therefore, in order to solve \eqref{mconstr}, it is enough to let $m$ be a multiple of $k/2$. Overall, we find that all the $\beta_{a,\ell=0,m}$ terms in \eqref{WTK1} are functions of $y_1^r$ and $y_2^r$. 

At this stage, the trinion SCFT can be found by setting to zero all the $\beta_{a,\ell,m}$ parameters with $\ell\neq0$ and define $w=y_1^r+1$. By properly tuning the remaining parameters we can bring \eqref{WTK1} to the form 
\be 
W_{T^2/\BZ_r}(x_1,x_2,x_3)+w^{p/r}=0\coma 
\ee 
which is precisely the hypersurface singularity describing the trinion SCFTs we are after with gauge group $\SU(p/r)=\SU(2n)$ when $k$ is even and $\SU(p/r)=\SU(n)$ when $k$ is odd. We would now like to notice that by restricting further the $\beta$ parameters, setting, e.g., to zero all $\beta_{a,\ell=0,m}$ with $m$ not a multiple of $k$, we can introduce a different variable $w=y_1^k+1$ and tune the surviving parameters in such a way that the resulting 3-fold reads 
\be \label{diffvar} 
W_{T^2/\BZ_r}(x_1,x_2,x_3)+w^{n}=0\fstop
\ee 
This leads to trinions of $D_p(\SU(n))$ theories also for $k$ even. The conclusion is that {\it $\SU(p/r)$ is the maximal choice}, but with a different specialization of the parameters, we can also find the same trinion theory but with {\it a $\SU(n')$ vector multiplet where $n'$ is a divisor of $p/r$}. All choices such that $n'$ is coprime with all the $p_i$'s labelling the $D_{p_i}(\SU(n'))$ sectors will lead to SCFTs with $a=c$ central charges (see \cite{Kang:2021lic, Buican:2020moo, Kang:2021lic, Kang:2022vab}).\footnote{The case of $k=8$ in \eref{eq:DpgaugingSCFTs} with $\SU(2n)$ replaced by $\SU(3)$ describes $(A_3, E_6)$, and the case of $k=12$ in \eref{eq:DpgaugingSCFTs} with $\SU(2n)$ replaced by $\SU(5)$ describes $(A_5, E_8)$; see also \cites[Appendix C]{Carta:2021whq}[Appendix C]{Carta:2022spy}.}\footnote{Note that, in the case of maximal choice, only the theories with $k=6$ have central charges $a=c$.} An analogous argument to the one above when $n$ is a multiple of $k$ leads to the Lagrangian SCFTs \eqref{eq:LagrangianSCFTs} (see \cite{DelZotto:2015rca} for the details), which should be seen as the maximal choice for this set of theories. 

Let us consider a couple of examples to illustrate the above arguments. For $k=6$ and $n=5$ the orbifold group is $\BZ_{30}$ and acts as follows: 
\be
\mathbb{Z}_6: \;(\omega_6,\omega_6;\omega_6^{-2})\,; \quad \mathbb{Z}_5: \; (\omega_5,\omega_5^{-1};1)\fstop 
\ee
We can find trinion SCFTs made of three $D_3(\SU(N))$ theories gauged together with $N=2,\,5$ or $10$, where $\SU(10)$ is the maximal choice. For $N=2$ we find the $(A_2,D_4)$ theory, which is located at the point $\beta_{\ell=0,m=15}=2$,\footnote{We have suppressed the parameter $a$ since we have only one chiral operator for $\ell=0$.} and all other $\beta$ parameters set to zero. From \eqref{contfrac}, we find that $p=30$ and $q=19$. The configuration of curves is, therefore, $(2,\,3,\,2,\,3,\,2)$ and after blowup we find (see \cite{Heckman:2013pva}) the sequence $(3,\,1,\,6,\,1,\,3,\,1,\,6,\,1,\,3,\,1,\,6,\,1,\,3)$ where the $-3$ curves support a $\mathfrak{su}(3)$ gauge algebra and the $-6$ curves support a $\mathfrak{e}_6$. The rank of the gauge algebra is therefore $26$ and the dimension of the tensor branch is $13$, leading to a total of $39$. This matches precisely the counting of $\beta$ parameters is the Landau-Ginzburg superpotential:
\be 
W_{\CT_{\text{KK}}} = W_{T^2/\BZ_3} + y_1^{30}+ y_2^{30} + \sum_a\sum_{\ell}\sum_m \beta_{a,\ell,m}\varphi_{a,\ell/3}(x_i)y_1^m y_2^{[19m]_{30}}\fstop
\ee
For $\ell=0$, we have one chiral operator ($\varphi(a,\ell/r=0)=1$) and $m$ is any multiple of $3$ ranging from $3$ to $27$. This provides nine parameters. For $\ell=1$, we have three chiral operators ($\varphi(a,\ell/r=1/3)=x_1,x_2,x_3$) and $m$ is of the form $3N+1$ with $0\leq N\leq 6$ giving a total of twenty-one $\beta$ parameters. Finally, for $\ell=2$, we have three chiral operators ($\varphi(a,\ell/r=2/3)=x_1x_2,x_2x_3,x_1x_3$) and $m=2,\,5$ or $8$ providing nine $\beta$ parameters. Overall, we reproduce the counting of thirty-nine parameters from 6d.

Let us also discuss the closely-related example with $k=3$ and $n=5$. In this case, the orbifold group is $\BZ_{15}$ and acts as 
\be
\mathbb{Z}_3: \;(\omega_3,\omega_3;\omega_3)\,; \quad \mathbb{Z}_5: \; (\omega_5,\omega_5^{-1};1)\fstop 
\ee 
The only trinion SCFT we find in this example is given by a $\SU(5)$ vector multiplet gauging three copies of $D_3(\SU(5))$:
\bes{
\begin{tikzpicture}[scale=0.86,font=\footnotesize,baseline=0.25cm]
\node (a1) at (0,0)    {$\SU(5)$};
\node (a3) at (0,1)    {$D_3(\SU(5))$};
\node (a4) at (-2.2,0) {$D_3(\SU(5))$};
\node (a6) at (2.2,0)  {$D_3(\SU(5))$};
\draw (a3)--(a1) (a4)--(a1) (a6)--(a1);
\end{tikzpicture}
}
Note that this theory has $a=c=16$. From \eqref{contfrac}, we get $p=15$ and $q=4$, leading to the sequence of curves $(4,\,4)$, which become after the required blowups $(6,\,1,\,3,\,1,\,6)$ which corresponds to the minimal $(E_6,E_6)$ conformal matter with both $\mathfrak{e}_6$ algebras gauged. We have a gauge group of rank 14 and 5 tensors, leading to a total of nineteen parameters. The LG superpotential reads 
\be 
W_{\CT_{\text{KK}}} = W_{T^2/\BZ_3} + y_1^{15}+ y_2^{15} + \sum_a\sum_{\ell}\sum_m \beta_{a,\ell,m}\varphi_{a,\ell/3}(x_i)y_1^m y_2^{[4m]_{15}}\fstop
\ee
From \eqref{mconstr} we find, for $\ell=0$, four $\beta$ parameters with $m=3,\,6,\,9,\,12$. For $\ell=1$, we get nine parameters with $m=2,\,5,\,8$, and six for $\ell=2$, corresponding to $m=1,\,4$. The total is indeed nineteen, as expected. 

\section{Non-invertible Symmetry in the \texorpdfstring{$(A_2, D_4)$}{} Theory}
\label{sec:A2D4}

In this section, we discuss in detail higher-form symmetries, their mixed anomalies, as well as the non-invertible symmetries of the $(A_2, D_4)$ theory. We will see that there is a close resemblance to those of the 4d $\CN=4$ super Yang-Mills theory with $\su(2)$ gauge algebra, and the arguments presented in \cite{Kaidi:2021xfk, Choi:2022zal, Kaidi:2022uux} can also be applied to this theory.

\subsection{Higher-form Symmetries}

As mentioned earlier, the $(A_2, D_4)$ Argyres-Douglas theory can be realized by taking three copies of the $(A_1, A_3) = D_3(\SU(2))$ theory, whose flavor symmetry algebra is $\su(2)$, and then gauging the diagonal $\su(2)$ symmetry \cite{Buican:2016arp, Closset:2020scj, Closset:2020afy}:
\bes{
\begin{tikzpicture}[scale=0.86,font=\footnotesize,baseline=0cm]
\node (a1) at (0,0)    {$\su(2)$};
\node (a3) at (0,1)    {$D_3(\SU(2))$};
\node (a4) at (-2.2,0) {$D_3(\SU(2))$};
\node (a6) at (2.2,0)  {$D_3(\SU(2))$};
\draw (a3)--(a1) (a4)--(a1) (a6)--(a1);
\end{tikzpicture}
}
Note that this theory has the central charges $a=c=2$. The corresponding global form of the gauge symmetry can be chosen to be $\SU(2)$, $\SO(3)_+$ or $\SO(3)_-$, giving rise to three variants of the $(A_2, D_4)$ theory, denoted by $(A_2, D_4)_0$, $(A_2, D_4)_+$ and $(A_2, D_4)_-$, respectively.  We will discuss the symmetries of the theory for each of these choices later.  In the meantime, let us discuss how to realize it from a higher dimensional perspective.

For completeness, let us summarize the procedure in obtaining the $(A_2, D_4)$ theory discussed in \cite{DelZotto:2015rca} and in the precedent section.  We first compactify F-theory on the Calabi-Yau 3-fold geometry $(T^2 \times \BC^2)/\BZ_6$. The resulting 6d $\cN=(1,0)$ theory is the minimal SCFT on the $(-6)$-curve. We then compactify this 6d SCFT further on $T^2_\tau$, where the subscript $\tau$ denotes the complex structure of this particular $T^2$. Keeping the size of the torus finite, we obtain a four dimensional KK theory $\CT_{\text{KK}}$, described by the Landau-Ginzburg mirror associated with $(T^2 \times \BC^2)/\BZ_6$ orbifold. The superpotential for the geometric engineering for $\CT_{\text{KK}}$ is given by \cite[(5.1)]{DelZotto:2015rca}
\bes{ \label{WTKK}
W_{\CT_{\text{KK}}} &= W_{T^2/\BZ_3}(\tau; x_1, x_2, x_3) + y_1^6+ y_2^6 + y_1 y_2(\beta_1 x_1 x_2 + \beta_2 x_1 x_3 + \beta_3 x_2 x_3) \\
& \quad +y_1^2 y_2^2 (\beta_4 x_1 + \beta_5 x_2 +\beta_6 x_3) + \beta_7 y_1^3 y_2^3\coma
}
where the part associated with the $T^2$ torus\footnote{This is actually the $T^2$ in the mirror geometry $(T^2 \times \BC^2)/\BZ_6$. It should not to be confused with $T^2_\tau$, which is involved in the torus compactification of the 6d theory to $\CT_{\text{KK}}$.} is described by three complex variables $x_1$, $x_2$ and $x_3$ such that
\bes{
W_{T^2/\BZ_3}(\tau; x_1, x_2, x_3) = x_1^3+x_2^3+x_3^3 + \tau x_1 x_2 x_3
}
and the $\BC^2$ part is described by two complex variables $y_1$ and $y_2$. Note that the complex structure $\tau$ of the $T^2_\tau$ torus appears as a dimensionless parameter in the superpotential. Subsequently, we go to a special point on the Coulomb branch of $\CT_{\text{KK}}$ where the $(A_2, D_4)$ theory emerges.  This is done by tuning the parameters involving only $y_1$ and $y_2$ in order to get a quadratic term, and this is performed in such a way that the $W_{T^2/\BZ_3}$ part is left untouched: 
\bes{
\tilde{W}_{\CT_{\text{KK}}} &= x_1^3+x_2^3+x_3^3 + \tau x_1 x_2 x_3+ (y_1^3+y_2^3)^2\coma
}
The singular geometry associated with the geometric engineering of the $(A_2, D_4)$ theory is obtained by setting $y_2=1$ and taking $w=y_1^3+1$:
\bes{ \label{eqA2D4}
 x_1^3+x_2^3+x_3^3 + \tau x_1 x_2 x_3 + w^2 =0~.
}
We will shortly see that the higher-form symmetry of the $\CT_{\text{KK}}$ theory is larger than that of the $(A_2, D_4)$ theory.  In other words, part of the 1-form symmetry of $\CT_{\text{KK}}$ acts trivially on the line operators of the $(A_2, D_4)$ theory.

The BPS strings associated with the $(-6)$-curve gives rise to a $\BZ_6^{[2]}$ 2-form symmetry \cite{DelZotto:2015isa,Apruzzi:2020zot, Bhardwaj:2020phs}. Moreover, there is also a $\mathfrak{e}_6$ gauge algebra associated with the $(-6)$-curve. The precise gauge group can be chosen as either $E_6$ or $E_6/\BZ_3$, where the former choice gives rise to a $\BZ_3^{[1]}$ 1-form  symmetry and the latter choice gives rise to a $\BZ_3^{[3]}$ 3-form symmetry (see also the discussion in \cite{Apruzzi:2022dlm}).  Note that gauging one of these symmetries leads to the other; in other words, they are dual symmetries in 6d.\footnote{In general, gauging a $\BZ_N^{[p]}$ $p$-form symmetry in $d$ spacetime dimensions leads to a dual $\BZ_N^{[d-p-2]}$ $(d-p-2)$-form symmetry.}

Let us first consider the $\BZ_6^{[2]}$ symmetry in 6d.  Upon compactification on a circle to 5d, the $\BZ_6^{[2]}$ 2-form symmetry leads to a $\BZ_6^{[2]}$ 2-form symmetry and a $\BZ_6^{[1]}$ 1-form symmetry \cite{Gaiotto:2014kfa}.  However, as pointed out in \cite[Section 3.3.1]{Bhardwaj:2020phs}, the 5d theory cannot possess both 1-form and 2-form symmetries originating from the 2-form symmetry of the 6d theory. In other words, one has to choose to keep either the 1-form or 2-form symmetry.  In fact, gauging one leads to the other, \ie~ they are dual symmetries in 5d. For definiteness, we choose to keep the $\BZ_6^{[1]}$ 1-form symmetry. Upon further compactification on a circle to 4d, the latter gives rise to a $\BZ_6^{[1]}$ 1-form symmetry and a $\BZ_6^{[0]}$ 0-form symmetry.

Now let us consider the $\BZ_3^{[1]}$ symmetry in 6d. Upon compactification on a circle to 5d, this leads to a 1-form $\BZ_3^{[1]}$ symmetry and a 0-form $\BZ_3^{[0]}$ symmetry. Upon further compactification on a circle to 4d, the $\BZ_3^{[1]}$ symmetry gives rise to a $\BZ_3^{[1]}$ 1-form symmetry and a $\BZ_3^{[0]}$ 0-form symmetry, whereas the 0-form $\BZ_3^{[0]}$ symmetry remains the same. In summary, we have a $\BZ_3^{[1]} \times \left(\BZ_3^{[0]} \right)^2$ symmetry in 4d, originating from the $\BZ_3^{[1]}$ symmetry in 6d.  

On the other hand, suppose that we consider the $\BZ_3^{[3]}$ symmetry in 6d, using the same arguments as before. Upon compactifying on a torus to 4d, we would have a $\BZ_3^{[3]} \times \BZ_3^{[2]} \times \BZ_3^{[2]} \times\BZ_3^{[1]}$. Since, if gauged,  $\BZ_3^{[3]}$ would not be dynamical in 4d, the dynamical symmetries are $\BZ_3^{[1]} \times \left(\BZ_3^{[0]} \right)^2$, as mentioned in the previous paragraph, where we have dualized $\BZ_3^{[2]}$  to $\BZ_3^{[0]}$ in 4d.

Let us turn back to the $(A_2, D_4)$ theory. According to \cite{DelZotto:2020esg,Closset:2020scj}, this theory has a $\BZ_2^{[1]}$ 1-form symmetry. Since $\BZ_2$ is not a subgroup of $\BZ_3$, it is clear that this $\BZ_2^{[1]}$ must be a subgroup of the $\BZ_6^{[1]}$ 1-form symmetry, which originates from the $\BZ_6^{[2]}$ 2-form symmetry in 6d.  The other $\BZ_3$ symmetries as well as the $\BZ_3^{[1]}$ commutant of $\BZ_2^{[1]}$ in $\BZ_6^{[1]}$ act trivially on the $(A_2, D_4)$ theory.  In other words, by going to a special point of the Coulomb branch of $\CT_{\text{KK}}$ where the $(A_2, D_4)$ theory emerges, only the $\BZ_2^{[1]}$ 1-form symmetry subgroup of $\BZ_6^{[1]}$ of the former acts non-trivially on the line operators of $(A_2, D_4)$. 

We would like to end this subsection with a brief comment. The fact that the 1-form symmetry of $(A_2,D_4)$ is smaller than the 1-form symmetry of the mother KK theory is a manifestation of a phenomenon that seems to be very generic for AD theories. Such a decrease in the 1-form symmetry often happens when the AD theory is realized at a specific point in the (extended) Coulomb branch of some asymptotically free theory. For example, it even happens for the most basic AD theory, $H_0 = (A_1, A_2)$ (which has a trivial 1-form symmetry) when compared to the asymptotically free pure $\SU(3)$ gauge theory (which has a $\mathbb{Z}_3$ 1-form symmetry) in which we can embed the $H_0$ theory.

\subsection{\texorpdfstring{$\SL(2,\BZ)$}{} Transformations and Non-invertible Defects of \texorpdfstring{$(A_2, D_4)$}{}}
\label{sec:non-inveA2D4}

Let us mention some properties of the $(A_2, D_4)$ theory.  Recall that the three variants of the $(A_2, D_4)$ theory, namely $(A_2, D_4)_0$, $(A_2, D_4)_+$ and $(A_2, D_4)_-$, can be realized from gauging the diagonal $\SU(2)$ or $\SO(3)_+$ or $\SO(3)_-$ flavor symmetry of three copies of the $D_3(\SU(2))$ theory \cite{Buican:2016arp, Closset:2020scj, Closset:2020afy}. Each of these variants has a $\BZ_2^{[1]}$ 1-form symmetry, and the $\CN=2$ preserving conformal manifold is one complex dimensional \cite{Carta:2021whq}, where the parametrization can be taken as the holomorphic coupling of the corresponding gauge group. This can indeed be identified as $\tau$, the complex structure of the $T^2_\tau$ torus, that appears also in \eref{eqA2D4}.
Note that, on the other hand, $D_3(\SU(2))=(A_1, A_3)$ does not have a conformal manifold \cite{Giacomelli:2020ryy} and does not have a 1-form symmetry \cite{Closset:2020scj, DelZotto:2020esg}. 

As pointed out in \cite[Section 5.1]{DelZotto:2015rca}, even though direct compactification of the 6d theory on $T^2_\tau$ leads to $\CT_{\text{KK}}$, the parameter $\tau$ of the conformal manifold and the $\SL(2,\BZ)$ transformation that acts on $\tau$ (as well as the local and extended operators) of the $(A_2, D_4)$ theory are inherited directly from the complex structure and mapping class group of the torus $T^2_\tau$ respectively.\footnote{The presence of the $\SL(2,\BZ)$ S-duality group, together with the action on the mass parameters, was studied in the special case of the $(A_3,A_3)$ theory at the level of the SW curve in \cite{Buican:2014hfa}.}  This is a consequence of the fact that, in tuning the parameters in $\CT_{\text{KK}}$ to reach the $(A_2, D_4)$ superconformal point, the parameter $\tau$ is left untouched.

At this point, we observe that the $(A_2, D_4)_0$, $(A_2, D_4)_+$ and $(A_2, D_4)_-$ theories can be analyzed in the same way as the 4d $\CN=4$ SYM with gauge groups $\SU(2)$, $\SO(3)_+$ and $\SO(3)_-$ respectively. In the subsequent analysis, we follow closely the arguments given by \cite{Kaidi:2022uux} (see also \cite{Kaidi:2021xfk, Choi:2022zal}).  The $\mathsf{S}$ and $\mathsf{T}$ transformations in $\SL(2,\BZ)$ acts on the parameter $\tau$ as $\tau \rightarrow -1/\tau$ and $\tau \rightarrow \tau+1$, respectively. Let $B$ be a 2-form background gauge field of the $\BZ_2^{[1]}$ 1-form symmetry. In the same way as in \cite{Gaiotto:2014kfa, Bhardwaj:2020ymp, Choi:2022zal, Kaidi:2021xfk, Kaidi:2022uux} (which generalized \cite{Witten:2003ya}), we introduce topological manipulations $\CS$ and $\CT$ as follows: Let $\CS$ to be gauging of the $\BZ_2^{[1]}$ 1-form symmetry, and $\CT$ to be stacking an invertible field theory 
\bes{ \label{counterterm}
\exp\left(\frac{1}{2} \pi i \int \CP(B) \right)\coma
} 
where $\CP(B)$ is the Pontryagin square of $B$. In fact, \eref{counterterm} can be viewed as the counterterm resulting from the $\mathsf{T}$ action which shifts the theta angle $\theta \rightarrow \theta+2 \pi$.  In fact, $\CS$ and $\CT$ generates the $\SL(2,\BZ_2)$ symmetry; they satisfy the relations $\CS^2 = (\CS \CT)^3=1$. Following \cite{Kaidi:2022uux}, we add an extra subscript $m$ taking a $\mod \, 2$ value to $(A_2, D_4)_x$ (with $x=0, +, -$) in such a way that $(A_2, D_4)_{x,m}$ means the $(A_2, D_4)_{x}$ theory without the counterterm for $m=0 \, \mod \, 2$ and with the counterterm for $m=1 \, \mod \, 2$.  By shifting the $\theta$-angle by $2 \pi$ while turning on the background field $B$, we see that $\mathsf{T}$ maps $(A_2,D_4)_{0,0}$ to $(A_2,D_4)_{0,1}$; see \cite[Eqs. (2.1)-(2.2)]{Kaidi:2022uux}. Similarly, $\CT$ exchanges $(A_2,D_4)_{0,m}$ with $(A_2,D_4)_{0,m+1}$. Under the $\CS$ action, we have\footnote{As in \cite{Kaidi:2022uux}, we suppress the overall normalization of the partition function.}
\bes{
Z_{(A_2, D_4)_{+,0}}[\tau, B] &= \sum_{b \in H^2(X, \BZ_2)} Z_{(A_2, D_4)_{0,0}}[\tau, b] \, \exp\left(i \pi \int b B \right) \coma\\
Z_{(A_2, D_4)_{-,0}}[\tau, B] &= \sum_{b \in H^2(X, \BZ_2)} Z_{(A_2, D_4)_{0,1}}[\tau, b] \exp\left(i \pi \int b B \right)\\
&= \sum_{b \in H^2(X, \BZ_2)} Z_{(A_2, D_4)_{0,0}}[\tau, b] \, \exp\left(i \frac{\pi}{2} \int \CP(B) \right) \exp\left(i \pi \int b B \right)~.
}
As explained in \cite{Kaidi:2022uux}, $\CS^2=1$ implies that $\CS$ interchanges $(A_2, D_4)_{0,0}$ with $(A_2, D_4)_{+,0}$, and $(A_2, D_4)_{0,1}$ with $(A_2, D_4)_{+,1}$, and $(\CS \CT)^3=1$ implies that we can map any $(A_2, D_4)_{x,m}$ theory back to itself by applying $(\CS \CT)^3$. Moreover, using the fact that the group $\SL(2,\BZ_2)$ commutes with $\SL(2,\BZ)$, we have
\bes{
&\CS \mathsf{T} (A_2,D_4)_{0,0} = \CS (A_2,D_4)_{0,1} = (A_2,D_4)_{-,0} 
= \mathsf{T}\CS  (A_2,D_4)_{0,0} =  \mathsf{T} (A_2,D_4)_{+,0}\coma
}
which means that $\mathsf{T}$ maps $(A_2,D_4)_{+,0}$ to $(A_2,D_4)_{-,0}$.  Using the fact that $(\mathsf{S} \mathsf{T})^3=1$ together with the fact that $\SL(2,\BZ_2)$ commutes with $\SL(2,\BZ)$, we arrive to the same diagram as \cite[Figure 5]{Kaidi:2022uux}, namely
\bes{
\begin{tikzpicture}[baseline,font=\footnotesize]
\node[draw=none] (01) at (-4,1.5)  {$(A_2,D_4)_{0,1}$};
\node[draw=none] (p1) at (0,1.5)   {$(A_2,D_4)_{+,1}$};
\node[draw=none] (m1) at (4,1.5)   {$(A_2,D_4)_{-,1}$};
\node[draw=none] (00) at (-4,-1.5) {$(A_2,D_4)_{0,0}$};
\node[draw=none] (p0) at (0,-1.5)  {$(A_2,D_4)_{+,0}$};
\node[draw=none] (m0) at (4,-1.5)  {$(A_2,D_4)_{-,0}$};
\draw[Triangle-Triangle,blue] (00) to [bend left]  node[midway,  left] {\blue $\mathsf{T}$} (01);
\draw[Triangle-Triangle,blue] (01) to [bend left]  node[midway, above] {\blue $\mathsf{S}$} (p1);
\draw[Triangle-Triangle,blue] (p1) to [bend left]  node[midway, above] {\blue $\mathsf{T}$} (m1);
\draw[Triangle-Triangle,blue] (m1) to [bend left]  node[midway, right] {\blue $\mathsf{S}$} (m0);
\draw[Triangle-Triangle,blue] (m0) to [bend left]  node[midway, below] {\blue $\mathsf{T}$} (p0);
\draw[Triangle-Triangle,blue] (p0) to [bend left]  node[midway, below] {\blue $\mathsf{S}$} (00);
\draw[Triangle-Triangle,red]  (00) to [bend right] node[midway, right] {\red $\mathcal{T}$} (01);
\draw[Triangle-Triangle,red]  (p1) to [bend right] node[midway, below] {\red $\mathcal{S}$} (m1);
\draw[Triangle-Triangle,red]  (m1) to [bend right] node[midway,  left] {\red $\mathcal{T}$} (m0);
\draw[Triangle-Triangle,red]  (p0) to [bend right] node[midway, above] {\red $\mathcal{S}$} (00);
\draw[Triangle-Triangle,red]  (01) to node[midway, above, near start]  {\red $\mathcal{S}$} (m0);
\draw[Triangle-Triangle,red]  (p1) to node[midway, left, near end]     {\red $\mathcal{T}$} (p0);
\end{tikzpicture}
}

The subsequent statements, again, follow in the same way as for the 4d $\CN=4$ SYM with $\su(2)$ gauge algebra \cite{Kaidi:2022uux}. At $\tau= i$, the operation $\CS \mathsf{S}$ maps the $(A_2, D_4)_{0,0}$ theory back to itself and is implemented by a {\it non-invertible} defect in the theory. More generally, at $\tau=i$, the $(A_2, D_4)_{0,m}$ and $(A_2, D_4)_{+,m}$ theories have non-invertible duality defects $\CT ^m \CS \mathsf{S} \CT^{-m}$, whereas the $(A_2, D_4)_{-,m}$ theories have invertible defects $\CT ^m \CT \mathsf{S} \CT^{-m}$.  The presence of $\CT$ in the latter even for the case of $m=0$ indicates that there is a {\it mixed anomaly} between $\mathsf{S}$ and the $\BZ_2^{[1]}$ 1-form symmetry in the $(A_2, D_4)_{-,m}$ theory at $\tau =i$.  On the other hand, at $\tau= e^{2\pi i/3}$, the $(A_2, D_4)_{0,m}$ and $(A_2, D_4)_{-,m}$ theories have {\it non-invertible} triality defects $\CT^m \CS \CT \mathsf{S} \mathsf{T} \CT^{-m}$, whereas the $(A_2, D_4)_{+,m}$ theories have {\it non-invertible} triality defects $\CT^m  \CT \CS \mathsf{S} \mathsf{T} \CT^{-m}$.

\subsection{Anomalies Involving the \texorpdfstring{$\BZ_6^{[1]}$}{} 1-form Symmetry of the KK Theory}
\label{sec:conjmixedanomaly}

From \eref{WTKK}, we see that the parameter $\tau$ enters the superpotential via the term $W_{T^2/\BZ_3}$ and that tuning the parameters in $\CT_{\text{KK}}$ to reach the $(A_2, D_4)$ SCFT does not affect the term $W_{T^2/\BZ_3}$ in any way. We can therefore restrict ourselves to a one dimensional submanifold of the parameter space parametrized by $\tau$, on which the $\SL(2,\BZ)$ group acts in a usual way. We can also focus on $\tau=i$, which is the fixed point of the $\mathsf{S}$ transformation of $\SL(2,\BZ)$. For simplicity, let us choose the configuration of the 6d theory in such a way that $\CT_{\text{KK}}$ contains the $(A_2, D_4)_{-,0}$ theory as a SCFT point on the Coulomb branch. At this point, it makes sense to ask whether there is a mixed anomaly between $\mathsf{S}$ and the $\BZ^{[1]}_6$ 1-form symmetry.

Unfortunately, we do not know an explicit description of the KK theory, and so we cannot produce an answer to this question in a direct way.
Nevertheless, we have seen that there is a mixed anomaly between $\mathsf{S}$ and a $\BZ^{[1]}_2$ subgroup of the $\BZ^{[1]}_6$ symmetry in the $(A_2, D_4)_{-,0}$ theory.  It is therefore expected that the answer to the precedent question is {\it positive}. More generally, we will show the following statement holds:\footnote{We emphasize that this statement is not in contradiction with the proposition of \cite[Section 5.2]{Bashmakov:2022jtl}, namely there is a mixed anomaly between the $\BZ_2^{[0]}$ 0-form symmetry associated with $\mathsf{S}$ at $\tau = i$ and the $\BZ_2^{[1]}$ subgroup of the $\BZ_{2n}^{[1]} \times \BZ_{2n}^{[1]}$ 1-form symmetry of the 4d $\CN=4$ SYM with gauge group $\SU(4n^2)/\BZ_{2n}$. We will elucidate this point for $n=3$ below.}\\

\noindent\shadowbox{
\begin{minipage}{0.9581\textwidth}
\begin{claim}\label{obssubgroup}
If a $\BZ^{[1]}_{N}$ 1-form symmetry has a mixed anomaly with a $\BZ^{[0]}_2$ 0-form symmetry, then a $\BZ^{[1]}_n$ subgroup of $\BZ^{[1]}_{N}$ (\ie~ $n$ divides $N$) has a mixed anomaly with the $\BZ^{[0]}_2$ 0-form symmetry if $n^2$ does not divide $N$.  
\end{claim}
\end{minipage}
}\\

This result is analogous to that presented in \cite[(2.31)]{Hsin:2018vcg}. We will provide an argument to justify Claim \ref{obssubgroup} for general $N$ and $n$ below.  For the problem at hand, we have $N=6$, $n=2$, and $\mathsf{S}$ being identified with the $\BZ^{[0]}_2$ 0-form symmetry. Therefore, we find that the presence of a mixed anomaly between $\mathsf{S}$ and the $\BZ^{[1]}_6$ 1-form symmetry in the KK theory is consistent with the fact that there is a mixed anomaly between $\mathsf{S}$ and the $\BZ^{[1]}_2$ subgroup of the $\BZ^{[1]}_6$ 1-form symmetry in the $(A_2, D_4)_{-,0}$ theory.  

Let us point out that our current problem is very similar to that of the 4d SYM theory with gauge group $(\SU(36)/\BZ_6)_0$, where the subscript $0$ denotes the discrete theta angle. This theory has a $\BZ_6^{[1]}$ electric 1-form symmetry and a $\BZ_6^{[1]}$ magnetic 1-form symmetry.  Moreover, it is easy to see that the set of the charges of line operators \cite[(2.8)]{Aharony:2013hda}
\bes{
\{(z_e, z_m) | (z_e, z_m) = (6 \mathfrak{e} , 6 \mathfrak{m}) ~\mod \, 36~,\, \mathfrak{e} \in \BZ, \, \mathfrak{m} \in \BZ \}
}
is invariant under the $\mathsf{S}$ action that maps $(z_e, z_m)$ to $(z_m, - z_e)$, and is also invariant under the $\mathsf{T}$ action that shifts the continuous theta angle by $2 \pi$ \cite[(2.9)]{Aharony:2013hda}, which maps $(z_e, z_m)$ to $(z_e+z_m, z_m)$.  To put it another way, the global form $(\SU(36)/\BZ_6)_0$ is invariant under $\mathsf{S}$ and $\mathsf{T}$ up to stacking invertible phases.  It follows that the self-duality defect at $\tau = i$ is invertible; in other words, the $\BZ_2^{[0]}$ 0-form symmetry associated with $\mathsf{S}$ has a mixed anomaly with a $\BZ^{[1]}_6$ subgroup of the $\BZ^{[1]}_6 \times \BZ^{[1]}_6$ 1-form symmetry.\footnote{Note that this is completely analogous to the discussion of the $\CN=4$ SYM with gauge group $(\SU(4)/\BZ_2)_+ \cong \SO(6)_+$, whose 1-form symmetry is $\BZ_2^{[1]} \times \BZ_2^{[1]}$.  This theory resides in its own $\SL(2,\BZ)$ orbit (see \cite[Figures 4 and 6]{Aharony:2013hda}), and is therefore invariant under $\mathsf{S}$ and $\mathsf{T}$ up to stacking invertible phases. As argued in \cite{Kaidi:2022uux}, at $\tau = i$, the self-duality defect is invertible, implying the mixed anomaly between $\mathsf{S}$ and a $\BZ_2^{[1]}$ subgroup of the 1-form symmetry. It is also argued in \cite{Kaidi:2022uux} that the non-invertible defects of the other variants of the $\so(6)$ gauge theory can all be related to the invertible defects of $\SO(6)_+$ by gauging $\BZ_2^{[1]}$ subgroup of its 1-form symmetry.}  It was argued in \cite[Section 5.2]{Bashmakov:2022jtl} that, for the $(\SU(36)/\BZ_6)_0$ SYM, the $\BZ_2^{[1]}$ subgroup of the $\BZ^{[1]}_6$ subgroup of the 1-form symmetry also has a mixed anomaly with the $\mathsf{S}$ symmetry.  This is indeed consistent with Claim \ref{obssubgroup}, since $2^2=4$ does not divide $6$.

\subsubsection*{Justification of Claim \ref{obssubgroup}}

Let us now justify Claim \ref{obssubgroup} using background fields.  Let $\CA_1$ be a 1-form background field for the $\BZ^{[0]}_2$ 0-form symmetry, which is $\mathsf{S}$ in our problem, and let $\CB_2$ be a 2-form background field for the $\ZZ^{[1]}_N$ 1-form symmetry.\footnote{Here the subscripts denote the form degrees.} We will assume that $N$ is even.  The mixed anomaly between $\BZ^{[0]}_2$ and $\ZZ^{[1]}_N$ is characterized by the 5d anomaly theory whose action is given by \cite[(C1)]{Kaidi:2021xfk},\footnote{In \cite[(C1)]{Kaidi:2021xfk}, $p$ was chosen to be $p=N/\GCD(2,N)$.} \cite[(2.1)]{Bashmakov:2022jtl}:\footnote{These anomalies have been computed also in earlier works, e.g. in \cite{Wan:2018bns}.}
\bes{ \label{action1}
    S &=  \frac{2\pi p}{N} \int_{\mathcal{M}_5} \CA_1\wedge \frac{\mathcal{P}(\CB_2)}{2} \,\, \longrightarrow \,\, \frac{p N}{2\pi^2} \int_{\mathcal{M}_5} A_1\wedge B_2 \wedge B_2 \coma
}
where $p$ is an integer. On the right of the arrow, we proceeded as in \cites[(E.1) and (E.2)]{Hsin:2018vcg} and \cite[Footnote 8]{Kaidi:2023maf} (see also \cite[(A.6)]{Choi:2022jqy}).  In particular, we treat $B_2$ as a flat $\U(1)$ 2-form gauge field whose holonomy is restricted to be $2\pi/N$ times an integer; this practically makes it a $\BZ_N^{[1]}$ gauge field, \ie~ $B_2 \rightarrow \frac{2 \pi}{N} \CB_2$.  Similarly, $A_1$ is treated as a flat $\U(1)$ 1-form gauge field whose holonomy is restricted to be $2\pi/2$ times an integer; this makes it effectively a $\BZ_2^{[0]}$ gauge field, \ie~ $A_1 \rightarrow \frac{2 \pi}{2} \CA_1$. We did not write explicitly the Lagrange multipliers that impose these restrictions.

To gauge a $\ZZ_{n}^{[1]}$ subgroup of the $\ZZ_N^{[1]}$ 1-form symmetry, we add a Lagrange multiplier $C_2$, which is a 2-form dynamical gauge field, to the above action in the following way:
\begin{equation}  \label{action2}
    S' = \frac{p N}{2\pi^2} \int_{\mathcal{M}_5}A_1\wedge B_2 \wedge B_2 + \frac{n}{2\pi } \int_{\mathcal{M}_5} B_2\wedge dC_2\fstop
\end{equation}
Using the equation of motion of $C_2$, we have
\begin{equation} \label{defl1}
    n B_2 = d \ell_1\coma \, \text{or} \quad  B_2 = \frac{ 1}{n} d \ell_1\coma
\end{equation}
where $\ell_1$ is a dynamical $\U(1)$ $1$-form gauge field. Using the equation of motion of $B_2$, we have
\begin{equation} \label{eqB2a}
\begin{split}
 0=   \frac{p N}{\pi^2}(A_1\wedge B_2) + \frac{n}{2 \pi} dC_2  
\end{split}
\end{equation}
Therefore, 
\begin{equation} \label{eqB2}
   \frac{n}{2\pi} B_2 \wedge dC_2 
   = - \frac{p N}{\pi^2} A_1 \wedge B_2 \wedge B_2 \fstop
\end{equation}
Substituting \eref{defl1} and \eref{eqB2} in the action \eref{action2}, we obtain
\bes{ \label{action3}
    S' &= \frac{p N}{2\pi^2}\int_{\mathcal{M}_5} A_1\wedge B_2 \wedge B_2 - \frac{p N}{\pi^2} \int_{\mathcal{M}_5} A_1\wedge B_2 \wedge B_2 \\
    &= -\frac{p N}{2\pi^2} \int_{\mathcal{M}_5} A_1\wedge B_2 \wedge B_2 \\
    &= -\frac{ pN}{n^2} \frac{1}{4\pi^2 }  \int_{\mathcal{M}_5} d \ell_0 \wedge d \ell_1 \wedge d \ell_1~
    \fstop
}
where in the last line we have used \eref{defl1} and written $A_1 = \frac{1}{2} d \ell_0$ such that $\ell_0$ is a dynamical $\U(1)$ $0$-form gauge field, where the factor $1/2$ is there to make the holonomy of the $A_1$ gauge field $2\pi/2$ times an integer.  

Recall that the 5d $\U(1)$ Chern-Simons term is $\frac{c}{24 \pi^2} a \wedge da \wedge da$, where gauge invariance restricts the coefficients $c$ to take values in $6\BZ$ on a generic five-manifold (see \cite{Witten:1996qb} and \cite[(2.5)]{Intriligator:1997pq}).  As a consequence, under gauge transformation $a \rightarrow a + d \lambda$, the integral $\frac{1}{4 \pi^2} \int_{\CM_5} d \lambda \wedge da \wedge da$ must be an integer multiple of $2\pi$.  Therefore, regardless of the value of $p$, we see that if $\frac{N}{n^2}$ is an integer, then \eref{action3} is an integer multiple of $2\pi$, \ie~ the anomaly action \eref{action3} is trivial for integral values of $\frac{N}{n^2}$.
On the other hand, if $\frac{N}{n^2}$ is not an integer, then there exists a  $p$ such that $\GCD(p, n)=1$ rendering action \eref{action3} not a multiple of $2 \pi$; in other words, there is a mixed anomaly between a $\BZ_{n}^{[1]}$ subgroup of the $\BZ_N^{[1]}$ 1-form symmetry and the $\BZ_2^{[0]}$ 0-form symmetry corresponding to $\mathsf{S}$ at $\tau =i$.

\section{Generalization}
\label{sec:generalization}

Having discussed in detail the properties of the $(A_2, D_4)$ theory, we now examine other theories in \eref{eq:LagrangianSCFTs} and \eref{eq:DpgaugingSCFTs}.  We will see that the resemblance of the $(A_2, D_4)$ theory and the $\CN=4$ SYM with $\su(2)$ gauge algebra is a special case of the general feature of theories with $a=c$. We will also generalize our argument to theories $a\neq c$ whose conformal manifolds generically have dimension larger than one. We point out that if we restrict ourselves to an appropriate one dimensional submanifold of the latter, we recover the observed resemblance to $\CN=4$ SYM again.  We first study the models which have a class $\CS$ realization and draw a general lesson from them.

\subsection{Lessons from S-duality of Trinion and Class \texorpdfstring{$\CS$}{} Theories} 
\label{sec:Sdualitytrionion}

In this section, we will check that trinion theories we have studied so far indeed have a $\SL(2,\BZ)$ S-duality and how it affects the global form of the gauge group using class $\CS$ technology. We will limit ourselves to the case $k=4$ in \eref{eq:LagrangianSCFTs} and \eref{eq:DpgaugingSCFTs}, where we have a $\SU(N)$ vector multiplet gauging four copies of $D_2(\SU(N))$ since this system is known to have a class $\CS$ description involving regular punctures only. We believe the statements which follow apply to other values of $k$ as well, even though we do not have an explicit derivation other than the construction from 6d we have discussed above. 

When $n$ is coprime with 4 (in other words $n$ is odd), we can have either a $\SU(N=2n)$ gauge group at the center, which corresponds to the maximal choice, or a $\SU(n)$ gauge group at the center. The latter has $a=c$ and behaves like $\CN=4$ SYM with the same gauge group, whereas the former has $a\neq c$ and the 1-form symmetry is a proper subgroup of the center. On the other hand, the case $\SU(N)$ with $N$ a multiple of 4 is recovered by taking $n$ to be a multiple of 4, so corresponds to the $T^2$ compactification of non-minimal $(\SO(8),\SO(8))$ conformal matter rather than a non-Higgsable theory. We would now like to understand how S-duality acts in these theories. 

We can now notice that the 4d SCFTs described above have a class $\CS$ realization with regular punctures only: The case $N=2m+1$ odd involves four identical $A_{2m}$ twisted punctures on the sphere, whereas for $N$ even we have eight punctures on the sphere, four twisted and four untwisted of type $A_{N-1}$. Let us start from the case $N$ odd, which is simpler to analyze, and then proceed to the case of $N$ even.

\subsubsection*{$N$ Odd}

Twisted $A_{2m}$ trinions were studied in \cite{Beem:2020pry} where it was observed that there is one trinion describing two decoupled copies of $D_2(\SU(2m+1))$. This involves two minimal twisted punctures (which carry no global symmetry) and a full untwisted puncture whose global symmetry is the diagonal combination of the global symmetry of the two decoupled SCFTs; see \cite[(6.4)]{Beem:2020pry}. In order to construct the theory we are interested in, we just need to glue together two such trinion obtaining a sphere with 4 twisted minimal punctures:
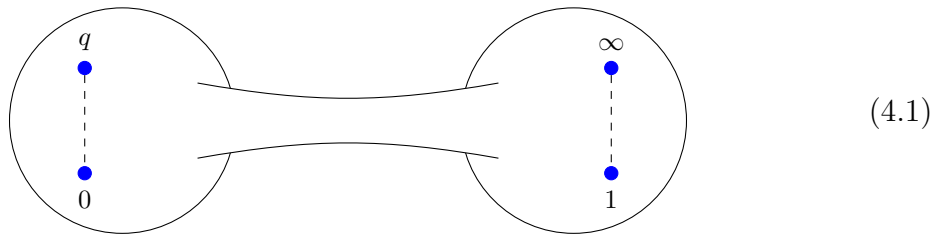
\be
\begin{tikzpicture}[baseline]
\tikzstyle{every node}=[font=\footnotesize]
\draw (-3,0) circle (1.5cm);
\draw (3,0) circle (1.5cm);
\node[circle,draw=blue,fill=blue,inner sep=0pt,minimum size=5pt,label={above:$q$}] (r1) at (-3.5,0.7) {};
\node[circle,draw=blue,fill=blue,inner sep=0pt,minimum size=5pt,label={below:$0$}] (r2) at (-3.5,-0.7) {};
\node[circle,draw=blue,fill=blue,inner sep=0pt,minimum size=5pt,label={above:$\infty$}] (r3) at (3.5,0.7)  {};
\node[circle,draw=blue,fill=blue,inner sep=0pt,minimum size=5pt,label={below:$1$}] (r4) at (3.5,-0.7)  {};
\coordinate (e) at (-2,0.5);
\coordinate (f) at (2,0.5);
\coordinate (g) at (-2,-0.5);
\coordinate (h) at (2,-0.5);
\fill[white] (e) to[bend right=10] (f) to (h) to[bend right=10] (g) to (e);
\draw[draw=black,solid,-]  (e) to[bend right=10]  (f) ;
\draw[draw=black,solid,-]  (g) to[bend left=10]  (h) ;
\draw[-,dashed] (r1)--(r2); 
\draw[-,dashed] (r3)--(r4);
\end{tikzpicture}
\ee 
As is well known, modulo reparametrizations of the sphere, we can set three punctures to be at $z=0$, $z=1$ and $z=\infty$. The position of the fourth cannot be constrained and parametrizes the conformal manifold (the parameter $q$ can be thought of as a function of the coupling $\tau$ we have introduced before). Let us now look at possible degeneration limits. When $q$ tends to zero, we have the weak-coupling limit, in which the two glued trinions disconnect and each one contains a pair of twisted punctures. 
When $q$ tends to either one or infinity, we find a degeneration limit, in which the two twist lines run along the tube. At this stage we can simply let them collide and reduce to the previous situation since their fusion leads to a trivial twist line. We display this for the case $q\rightarrow\infty$ in \eqref{cutrec}: 
\be
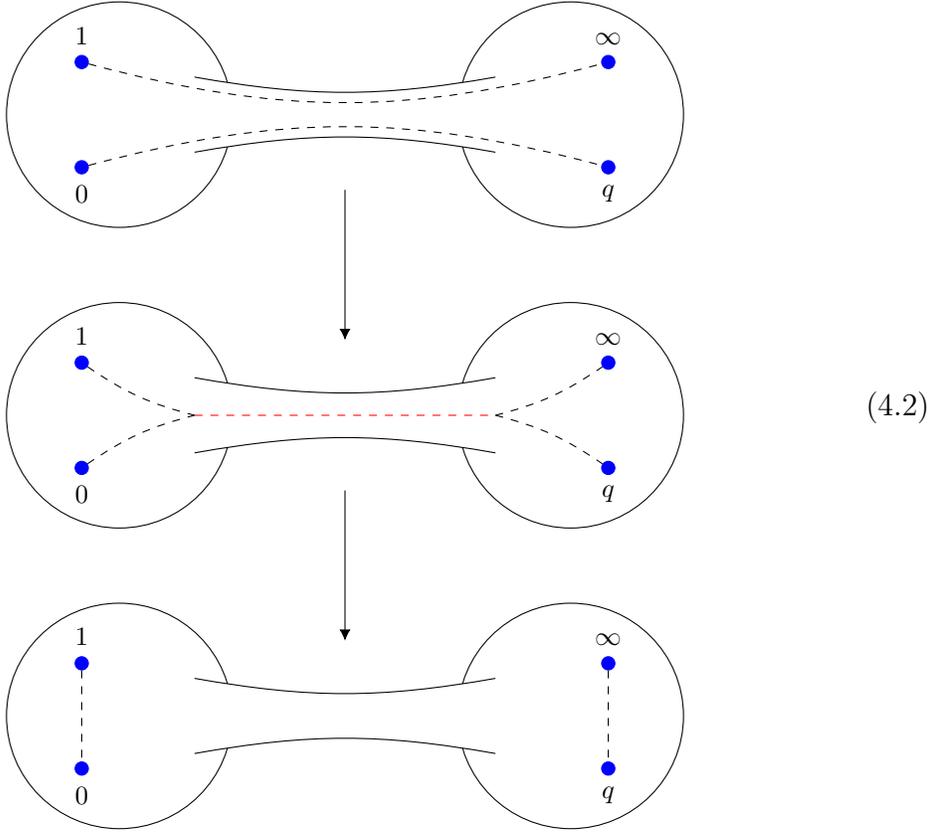
\label{cutrec}
\begin{tikzpicture}[baseline]
\tikzstyle{every node}=[font=\footnotesize]
\draw (-3,4) circle (1.5cm);
\draw (3,4) circle (1.5cm);
\node[circle,draw=blue,fill=blue,inner sep=0pt,minimum size=5pt,label={above:$1$}] (q1) at (-3.5,4.7) {};
\node[circle,draw=blue,fill=blue,inner sep=0pt,minimum size=5pt,label={below:$0$}] (q2) at (-3.5,3.3) {};
\node[circle,draw=blue,fill=blue,inner sep=0pt,minimum size=5pt,label={above:$\infty$}] (q3) at (3.5,4.7) {};
\node[circle,draw=blue,fill=blue,inner sep=0pt,minimum size=5pt,label={below:$q$}] (q4) at (3.5,3.3) {};
\coordinate (a) at (-2,4.5);
\coordinate (b) at (2,4.5);
\coordinate (c) at (-2,3.5);
\coordinate (d) at (2,3.5);
\fill[white] (a) to[bend right=10] (b) to (d) to[bend right=10] (c) to (a);
\draw[draw=black,solid,-]  (a) to[bend right=10]  (b) ;
\draw[draw=black,solid,-]  (c) to[bend left=10]  (d) ;
\draw[-,dashed] (q1) to[bend right=15] (q3); 
\draw[-,dashed] (q2) to[bend left=15] (q4);

\draw[->] (0,3)--(0,1);

\draw (-3,0) circle (1.5cm);
\draw (3,0) circle (1.5cm);
\node[circle,draw=blue,fill=blue,inner sep=0pt,minimum size=5pt,label={above:$1$}] (p1) at (-3.5,0.7)  {};
\node[circle,draw=blue,fill=blue,inner sep=0pt,minimum size=5pt,label={below:$0$}] (p2) at (-3.5,-0.7) {};
\node[circle,draw=blue,fill=blue,inner sep=0pt,minimum size=5pt,label={above:$\infty$}] (p3) at (3.5,0.7)   {};
\node[circle,draw=blue,fill=blue,inner sep=0pt,minimum size=5pt,label={below:$q$}] (p4) at (3.5,-0.7)  {};
\node[draw=none] (P) at (-2,0) {};
\node[draw=none] (Q) at (2,0) {};
\coordinate (A) at (-2,0.5);
\coordinate (B) at (2,0.5);
\coordinate (C) at (-2,-0.5);
\coordinate (D) at (2,-0.5);
\fill[white] (A) to[bend right=10] (B) to (D) to[bend right=10] (C) to (A);
\draw[draw=black,solid,-]  (A) to[bend right=10]  (B) ;
\draw[draw=black,solid,-]  (C) to[bend left=10]  (D) ;
\draw[-,dashed] (p1) to[bend right=10] (-2,0) (2,0) to[bend right=10] (p3); 
\draw[-,dashed] (p2) to[bend left=10] (-2,0) (2,0) to[bend left=10] (p4);
\draw[dashed,red] (-2,0)--(2,0);
\draw[->] (0,-1)--(0,-3);

\draw (-3,-4) circle (1.5cm);
\draw (3,-4) circle (1.5cm);
\node[circle,draw=blue,fill=blue,inner sep=0pt,minimum size=5pt,label={above:$1$}] (r1) at (-3.5,-3.3) {};
\node[circle,draw=blue,fill=blue,inner sep=0pt,minimum size=5pt,label={below:$0$}] (r2) at (-3.5,-4.7) {};
\node[circle,draw=blue,fill=blue,inner sep=0pt,minimum size=5pt,label={above:$\infty$}] (r3) at (3.5,-3.3)  {};
\node[circle,draw=blue,fill=blue,inner sep=0pt,minimum size=5pt,label={below:$q$}] (r4) at (3.5,-4.7)  {};
\coordinate (e) at (-2,-3.5);
\coordinate (f) at (2,-3.5);
\coordinate (g) at (-2,-4.5);
\coordinate (h) at (2,-4.5);
\fill[white] (e) to[bend right=10] (f) to (h) to[bend right=10] (g) to (e);
\draw[draw=black,solid,-]  (e) to[bend right=10]  (f) ;
\draw[draw=black,solid,-]  (g) to[bend left=10]  (h) ;
\draw[-,dashed] (r1)--(r2); 
\draw[-,dashed] (r3)--(r4);
\end{tikzpicture}
\ee 
We therefore see that all the degeneration limits are equivalent as in the case of four identical untwisted punctures. In that case it is known that the S-duality group of the theory is $\SL(2,\bZ)$, in perfect agreement with our expectations. We are also in the position to understand the action of the $\mathsf{S}$ transformation on the global structure of the gauge group. As was discussed in \cite{Bhardwaj:2021pfz}, the defect group of our theory is $\bZ_{N}^2$ and is generated by surface operators of the 6d SCFT wrapping a nontrivial cycle on the Riemann surface. The relevant cycles are those looping around one of the cuts, say the one connecting the punctures at $z=0,\,q$, and the one going from one cut to the other and then back, i.e.
\be
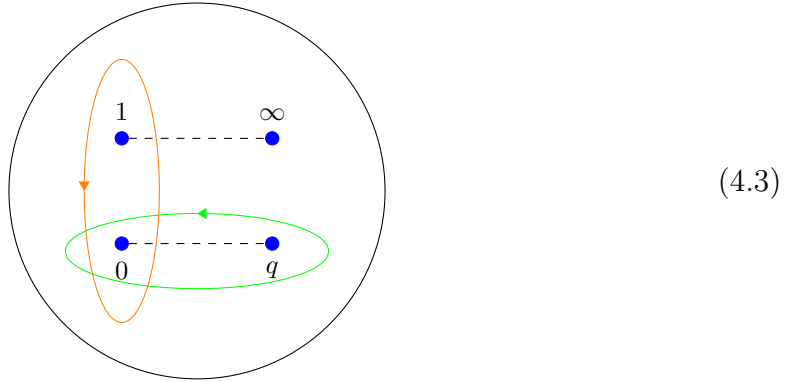
\begin{tikzpicture}[baseline=0cm]
\tikzstyle{every node}=[font=\footnotesize]
\draw (-2.75,0) circle (2.5cm);
\node[circle,draw=blue,fill=blue,inner sep=0pt,minimum size=5pt,label={above:$1$}] (q1)      at (-3.75,0.7) {};
\node[circle,draw=blue,fill=blue,inner sep=0pt,minimum size=5pt,label={below:$0$}] (q2)      at (-3.75,-0.7) {};
\node[circle,draw=blue,fill=blue,inner sep=0pt,minimum size=5pt,label={above:$\infty$}] (q3) at (-1.75,0.7) {};
\node[circle,draw=blue,fill=blue,inner sep=0pt,minimum size=5pt,label={below:$q$}] (q4)      at (-1.75,-0.7) {};
\draw[-,dashed] (q1) to[bend right=0] (q3); 
\draw[-,dashed] (q2) to[bend right=0] (q4);
\draw[decoration={markings,
    mark=at position {0.25} with {\arrow{Triangle}}},
    postaction={decorate},green] (-2.75,-0.8) ellipse (1.75cm and 0.5cm);
    \draw[decoration={markings,
    mark=at position {0.5} with {\arrow{Triangle}}},
    postaction={decorate},orange] (-3.75,0) ellipse (0.5cm and 1.75cm);
\end{tikzpicture}
\ee

It is easy to see that under the rearrangement of the twist lines we considered in \eqref{cutrec} these two cycles get interchanged and as a result electric and magnetic 1-form symmetries get interchanged as well. Therefore if we start from a theory with $\SU(N)$ gauge group (upon properly choosing the maximal subgroup of mutually local line operators) and $\bZ_N^{[1]}$ electric 1-form symmetry which becomes weakly coupled as $q\rightarrow 0$, when we send $q\rightarrow\infty$ and rearrange the twist lines we find a new weakly-coupled description. This will have a $\bZ_N^{[1]}$ magnetic 1-form symmetry and the gauge group is now $\SU(N)/\bZ_N$. This reproduces the structure of $\mathcal{N}=4$ SYM with $\su(N)$ gauge algebra. 

\subsubsection*{$N$ Even}

Let us now consider the case $N=2m$ even. Now, our trinion theory is Lagrangian and has a five dimensional conformal manifold. This model has a class $\CS$ realization in twisted $A_{2m-1}$ involving a sphere with 4 minimal untwisted and 4 minimal twisted punctures \cite{Chacaltana:2012ch}: 
\be
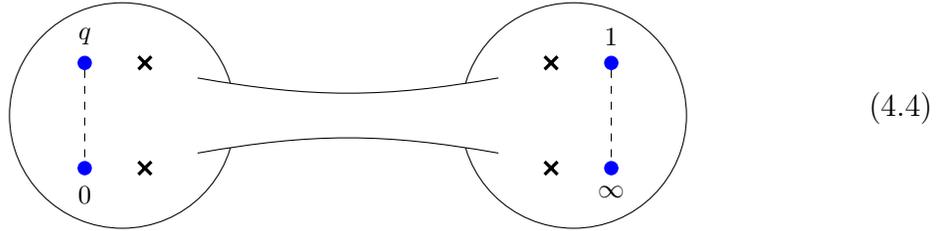
\label{oddsphere}
\begin{tikzpicture}[baseline]
\tikzstyle{every node}=[font=\footnotesize]
\draw (-3,0) circle (1.5cm);
\draw (3,0) circle (1.5cm);
\node[circle,draw=blue,fill=blue,inner sep=0pt,minimum size=5pt,label={above:$q$}] (p1) at (-3.5,0.7) {};
\node[circle,draw=blue,fill=blue,inner sep=0pt,minimum size=5pt,label={below:$0$}] (p2) at (-3.5,-0.7) {};
\node[circle,draw=blue,fill=blue,inner sep=0pt,minimum size=5pt,label={above:$1$}] (p3) at (3.5,0.7) {};
\node[circle,draw=blue,fill=blue,inner sep=0pt,minimum size=5pt,label={below:$\infty$}] (p4) at (3.5,-0.7) {};
\node[cross=3.7pt,line width=1.25pt,black] at (-2.7,0.7) {};
\node[cross=3.7pt,line width=1.25pt,black] at (-2.7,-0.7)   {};
\node[cross=3.7pt,line width=1.25pt,black] at (2.7,0.7)   {};
\node[cross=3.7pt,line width=1.25pt,black] at (2.7,-0.7)   {};
\coordinate (A) at (-2,0.5);
\coordinate (B) at (2,0.5);
\coordinate (C) at (-2,-0.5);
\coordinate (D) at (2,-0.5);
\fill[white] (A) to[bend right=10] (B) to (D) to[bend right=10] (C) to (A);
\draw[draw=black,solid,-]  (A) to[bend right=10]  (B) ;
\draw[draw=black,solid,-]  (C) to[bend left=10]  (D) ;
\draw[-,dashed] (p1)--(p2); 
\draw[-,dashed] (p3)--(p4);
\end{tikzpicture}
\ee 
We can now proceed as in the previous case and rearrange the twist lines as we go from one degeneration limit to another since they are not affected by the presence of untwisted punctures. The only caveat is that with this procedure we always land on equivalent cusps provided that we move the untwisted punctures accordingly, in such a way that in the degeneration limit we end up with two spheres, each with two twisted and two untwisted punctures. This clearly illustrates the following fact: The $\SL(2,\bZ)$ duality group we consider refers to a one dimensional submanifold of the conformal manifold. Only all cusps inside this special submanifold are equivalent. Said differently, the actual duality group of the theory is larger, our $\SL(2,\bZ)$ is a proper subgroup and its action maps the submanifold parametrized by $q$ to itself. We stress that we are not claiming that all the cusps in the full conformal manifold are equivalent, which is actually known not to be true.\footnote{For instance, the dualities for trinion theories studied in \cite{Closset:2020afy} provide plenty of counterexamples.} 

According to \cite{Bhardwaj:2021pfz} the defect group of \eqref{oddsphere} is $\bZ_m^2$ and therefore if we start from a $\SU(2m)$ gauge group in the center with electric $\bZ_m^{[1]}$ 1-form symmetry we end up, after rearranging the cuts, with a $\SU(2m)/\bZ_m$ gauge group and magnetic $\bZ_m^{[1]}$ 1-form symmetry. This is how the $\mathsf{S}$ transformation acts on the global structure of the gauge group. This is equivalent to the action of S-duality for $\mathcal{N}=4$ SYM with $\su(m)$ gauge algebra. 

We conclude by noticing that, although the action of S-duality in the examples we have studied depends only on the charge of the matter sector under the center of the gauge group, this is not true in general. For example, a trinion theory with a $\SU(4)$ gauge group at the center behaves like $\mathcal{N}=4$ $\SU(2)$ SYM whereas $\SU(4)$ SYM with one hypermultiplet in the rank-2 symmetric representation and one in the rank-2 antisymmetric representation has a different S-duality group. As studied in \cite{Chacaltana:2012ch}, the latter model corresponds in class $\CS$ of type $A_3$ to a torus with one untwisted puncture and a closed twist line. After a S-duality transformation the twist line ends up looping around the other cycle of the torus and as a result we get a $\USp(4)$ vector multiplet coupled to the $E_6$ Minahan-Nemeschansky theory \cite{Minahan:1996cj}. This cusp in the conformal manifold is clearly different from the one at which the $\SU(4)$ group becomes weakly-coupled.

\subsection{Theories in \eref{eq:LagrangianSCFTs} and \eref{eq:DpgaugingSCFTs}}

We now examine the non-invertible symmetries in general theories \eref{eq:LagrangianSCFTs} and \eref{eq:DpgaugingSCFTs}.  It is convenient to discuss two separate cases, namely theories with central charges $a=c$ and those with $a \neq c$.

\subsubsection*{Theories with $a=c$}

The theories with $a=c$ are those listed in \eref{eq:DpgaugingSCFTs} with the following conditions (see also \cite{Buican:2020moo, Kang:2021lic, Kang:2022vab}):
\ben
\item those with $k=3$ and $\GCD(n,3)=1$,
\item those with $k=6$ and $\GCD(n,6)=1$, and
\item the non-maximal choices discussed around \eref{diffvar}, where the central gauge group is $\SU(n')$ such that $n'$ is coprime with all the $p_i$'s labelling the $D_{p_i}(\SU(n'))$ sectors.
\een
 Each of these theories has a one dimensional conformal manifold parametrized by $\tau$, which can be viewed as the complex structure of the torus associated with compactification from 6d as well as the holomorphic coupling of the central gauge group.  The 1-form symmetry in each case is $\BZ^{[1]}_n$, $\BZ^{[1]}_{2n}$ and $\BZ^{[1]}_{n'}$, respectively. The action of the $\SL(2,\BZ)$ duality group on $\tau$, as well as the topological manipulations involving gauging of 1-form symmetries and stacking invertible phases are the same as for the 4d $\CN=4$ SYM with $\su(n)$, $\su(2n)$ and $\su(n')$ gauge algebras, respectively \cite{Kaidi:2021xfk, Choi:2022zal, Kaidi:2022uux}.  For appropriate values of $n$ and $n'$, there can be non-invertible duality and triality defects, as for the $\CN=4$ SYM.

\subsubsection*{Theories with $a\neq c$}

Theories with $a\neq c$ are 
\ben
\item \label{class1} those in \eref{eq:DpgaugingSCFTs} with $k=2 M$ even, $k\neq 6$, and the central gauge group being $\SU(2n)$ such that $\GCD(n, k)=\GCD(n, 2M)=1$, and 
\item \label{class2} those with the Lagrangian descriptions in \eref{eq:LagrangianSCFTs}.
\een
In general, the dimension of the conformal manifolds of these theories is larger than 1.  We restrict ourselves to a one dimensional submanifold of the conformal manifold parametrized by $\tau$, corresponding to the holomorphic gauge coupling of the central node.  The $\SL(2,\BZ)$ proper subgroup of the S-duality group acts on $\tau$ in a usual way. Let us work with an appropriate value of $n$ such that there is a global form of the central gauge group that is invariant under an element of $\SL(2,\BZ)$.  For simplicity, we take it to be $\mathsf{S}$. Focusing on $\tau =i $, we then obtain a 0-form symmetry of the theory.  

For theories in Class \ref{class1}, each of them has a $\BZ_n^{[1]}$ 1-form symmetry, whereas the 2-form symmetry of the corresponding parent 6d theory is $\BZ_{2Mn}^{[2]} \cong \BZ_{2M}^{[2]} \times \BZ_n^{[2]}$. As argued earlier, the KK theory also has a $\BZ_{2Mn}^{[2]}$ 1-form symmetry, where only the $\BZ_n^{[1]}$ subgroup acts non-trivially on the line operators of the corresponding  SCFT.  We can then proceed as discussed around Claim \ref{obssubgroup}. If the $\BZ_{2Mn}^{[1]}$ 1-form symmetry of the KK theory has a mixed anomaly with the 0-form symmetry $\mathsf{S}$ and $n^2$ does not divide $2Mn$, it follows that the $\BZ_n^{[1]}$ subgroup of the $\BZ_{2M}^{[1]} \times \BZ_n^{[1]}$ 1-form symmetry also has a mixed anomaly with the $\mathsf{S}$ 0-form symmetry. Gauging the $\BZ_n^{[1]}$ 1-form symmetry leads to the non-invertible symmetry \cite{Kaidi:2021xfk}.   

For theories in Class \ref{class2}, the situation is even simpler, since the 1-form symmetry of the 4d SCFT is the same as that of the KK theory and is equal to the 2-form symmetry of the parent 6d theory. We can thus choose a global form of the theory such that the 1-form symmetry has a mixed anomaly with the 0-form symmetry $\mathsf{S}$, and by gauging the former we obtain the non-invertible symmetry.

Let us consider this in an explicit example of $k=4$ and $n=8 r^2$ in \eref{eq:LagrangianSCFTs}. This also admits the class $\CS$ description \eref{oddsphere} with $m=4r^2$.  The original electric 1-form symmetry of the theory as depicted in \eref{eq:LagrangianSCFTs} is $\BZ_{4 r^2}$, but we can choose the global form of the gauge group such that it becomes $\BZ_{2r}^{[1]} \times \BZ_{2r}^{[1]}$, where one $\BZ_{2r}^{[1]}$ factor is the electric 1-form symmetry and the other is the magnetic 1-form symmetry. Focusing on the one-dimension submanifold of the conformal manifold parametrized by $\tau$, we find that the property of the theory is similar to that of the $\CN=4$ SYM with gauge group $(\SU(4r^2)/\BZ_{2r})_0$, which resides in its own $\SL(2,\BZ)$ orbit.  At $\tau=i$, the $\mathsf{S}$ generator of $\SL(2,\BZ)$ gives rise to a $\BZ_2^{[0]}$ 0-form symmetry. In the same way as argued in \cite[Section 5.2]{Bashmakov:2022jtl}, there is a mixed anomaly between the 0-form symmetry $\mathsf{S}$ and the $\BZ_2^{[1]}$ subgroup of the 1-form symmetry.  Gauging the latter leads to a non-invertible duality defect in the theory in question.

\section{Conclusions and Outlook}

In this paper, we have studied the presence of non-invertible symmetries in a certain class of Argyres-Douglas theories. In particular, we considered AD theories obtained by compactifying certain 6d $\mathcal{N}=(1,0)$ theories on a 2-torus $T^2$.  Such 6d theories can be constructed by compactifying F-theory on the orbifold $(\BC^2\times T^2)/\Gamma$, where $\Gamma$ are some cyclic subgroups of $\SU(3)$. By tuning the parameters of the theory, one can find a 4d $\mathcal{N}=2$ SCFT on the singular orbifold points. The resulting theories are then ``trinions'' arising from diagonal gauging of the $\su(N)$ flavor algebra of the $D_{p_i}(\SU(N))$ theories for values of $p_i$ shown in \eqref{tablep}. Various global forms of $\su(N)$ can be chosen in such a way that it is compatible with the operators of the $D_{p_i}(\SU(N))$ theories.  The compactification on torus with complex structure $\tau$ gives rise to a $\SL(2,\ZZ)$ group with $\mathsf{S}$ and $\mathsf{T}$ generators acting on $\tau$ as $\tau \rightarrow -1/\tau$ and $\tau \rightarrow \tau+1$, respectively. The parameter $\tau$ is identified with the holomorphic gauge coupling of the central gauge algebra $\su(N)$ in the trinion. The 1-form symmetry of these 4d theories are subgroups of the 2-form symmetry of their parent 6d theories. 

For the theories with central charges $a=c$, the conformal manifold is one dimensional, parametrized by $\tau$, and the discussion is parallel to that of $\CN=4$ SYM with gauge algebra $\fg = \su(N)$.  In particular, by choosing an appropriate global form of $\su(N)$, there are non-invertible duality and triality defects in those AD theories, in analogy with the $\CN=4$ SYM theory \cite{Kaidi:2021xfk, Choi:2021kmx, Choi:2022zal, Kaidi:2022uux}.  We have explicitly discussed this in the example of the $(A_2, D_4)$ theory.  It was observed in \cite{Buican:2020moo, Kang:2021lic} that the Schur indices for this class of theories can be written in terms of that of $\CN = 4$ SYM theory upon rescaling of fugacities (see also the comment in \cite{Kimura:2022yua}). It would be nice to investigate in future work whether these connections between the $\CN=2$ SCFTs with $a=c$ and the $\CN=4$ SYM are actually related or not.

For the theories with central charges $a \neq c$, the dimension of the conformal manifold is generally larger than one, and we restrict ourselves to the one dimensional subspace parametrized by $\tau$. In this class of theories, the center symmetry $\BZ_N$ of the central gauge group $\SU(N)$ is screened by certain operators of the surrounding $D_{p_i}(\SU(N))$ theories, and so the 1-form symmetry is a proper subgroup $\BZ_s$ of $\BZ_N$. The discussion is analogous to that of the $\CN=4$ SYM with $\su(s)$ gauge algebra. We demonstrated this observation by exploiting a class $\CS$ realization of a subclass of these theories.

Our findings lead to a number of open questions that are worth further investigations in the future. Let us list some of them as follows. First, it is natural to ask whether there are non-invertible symmetries in other families of $\CN=2$ AD theories or not. If so, it would be nice to examine the common properties of such theories and the dynamical consequences of the non-invertible symmetries.   Secondly, since there are several known 4d $\CN=1$ SCFTs with central charges $a=c$ that arise from diagonally gauging the $G$ symmetry of some $D_{p_i}(G)$ theories \cite{Kang:2021ccs, Kang:2023pot}, it is interesting to study the presence of the duality and triality defects in such theories.   Moreover, recently, it was discovered that some of such $\CN=1$ theories flow to a fixed point on the $\CN=1$ preserving conformal manifold of $\CN=4$ SYM \cite{Kang:2023pot}.  This could be a signal for the non-invertible symmetry in the aforementioned class of theories.

\acknowledgments 

We thank Christian Copetti, Michele Del Zotto and Craig Lawrie for useful discussions.  F. C. is supported by STFC consolidated grant ST/T000708/1. The work of S. G. is supported by the INFN grant ``Per attività di formazione per sostenere progetti di ricerca'' (GRANT 73/STRONGQFT). N. M. thanks the visiting research fellowship of the CNRS and the LPTENS, ENS Paris, where part of this project was conducted. A. M. is supported in part by Deutsche Forschungsgemeinschaft under Germany's Excellence Strategy EXC 2121  Quantum Universe 390833306. 

\appendix

\section{1-form Symmetries for Trinion Theories from 6d}
\label{sec:1formtrinion}

In this appendix, we discuss the 1-form symmetries for theories in \eqref{eq:DpgaugingSCFTs}. In order to do so, it is important to understand how operators in the $D_p(\SU(N))$ theories that appear in the quivers transform under the center of the $\SU(N)$ gauge group, where $N=n$ for $k=3$ and $N=2n$ for $k=4,\, 6, \, 8, \, 12$. Since in general $D_p(\SU(N))$ theories do not admit a 4d $\CN=2$ Lagrangian description, it is convenient to exploit the fact that upon compactification to 3d the resulting theories can be described in terms of 3d $\CN=4$ gauge theories \cite[Eqs. (5.10)-(5.11)]{Giacomelli:2020ryy}. We also rely on the observation that the Higgs branch of the original 4d theory coincides with that of the corresponding 3d theory upon compactification.  

Using the notation introduced in \cite[(5.1)]{Giacomelli:2020ryy}, we define
\begin{equation}\label{eq:definitions}
    x = \left\lfloor \frac{N}{p}\right\rfloor \coma M = N-(x+1) \coma m = \GCD(p,N) \coma n=\frac{N}{m}\coma q = \frac{p}{m}\coma
\end{equation}
and we obtain the following quivers for 3d reduction of $D_p(\SU(N))$ theories involved in \eqref{eq:DpgaugingSCFTs} (see \cite[Eqs. (5.10)-(5.11)]{Giacomelli:2020ryy})
\begin{subequations}
\begin{align}
    (D_2(\SU(2n)))_{\text{3d}}\,: &\, [2n] - \SU(n)\coma \label{eq:3dreduD2SU2n}\\
    (D_3(\SU(N)))_{\text{3d}}\,: &\, [N] - \U(M) - \U(x)\coma\label{eq:3dreduD3SUN}\\
    (D_4(\SU(2n)))_{\text{3d}}\,: &\, [2n] - \U(M) - \SU(n) - \U(x)\coma\label{eq:3dreduD4SU2n}\\
    (D_6(\SU(2n)))_{\text{3d}} \,: &\, [2n] - \U(M) - \U(n+x) - \SU(n) - \U(n-x-1) - \U(x)\fstop\label{eq:3dreduD6SU2n}
\end{align}
\end{subequations}
The Higgs branch of each of these $D_p(\SU(N))$ theories contains the meson (moment map) transforming in the adjoint representation of the flavor symmetry algebra $\su(N)$.  In addition, when there is a $\SU(n)$ gauge group present in the quiver, there are also baryonic operators in the rank-$n$ antisymmetric representation $\wedge^n$ of the flavor symmetry algebra $\su(2n)$.  These can be constructed using the chiral fields in the hypermultiplets between $[2n]$ and $\SU(n)$ where the $\SU(n)$ gauge indices are contracted with the Levi-Civita tensor.\footnote{For \cref{eq:3dreduD4SU2n,eq:3dreduD6SU2n} we can notice that it is not possible to build more baryonic-like operators using the bifundamental fields on the right-hand side of $\SU(n)$ since there is no way to make them gauge invariant.}  

When the $\SU(N)$ symmetries of these $D_p(\SU(N))$ theories are diagonally gauged as in \eqref{eq:DpgaugingSCFTs}, we have to check whether the Wilson lines in the representations charged under the $\mathbb{Z}_N$ center of the gauge group $\SU(N)$ may be screened or not by the Higgs branch operators. The mesons of course do not screen these Wilson lines.  However, for $N=2n$, the baryons in the representation $\wedge^n$ of $\su(2n)$ can screen them modulo $n$, and therefore the 1-form symmetry of the theory is reduced to $\BZ_n$.  Note that the $D_p(\SU(N))$ theory does not have a 1-form symmetry \cite{Hosseini:2021ged} (see also \cite{Closset:2020scj,DelZotto:2020esg, Closset:2021lwy}) and so there is no extra contribution to the 1-form symmetry of the star-shaped quivers \eqref{eq:DpgaugingSCFTs}.

In conclusion, we have the following results for \eqref{eq:DpgaugingSCFTs}:
\bi
\item There is no baryon in the $D_3(\SU(N))$ theory and so the 1-form symmetry for the $k=3$ case is $\BZ_n$, while for $k=6$ it is $\BZ_{2n}$.
\item For $k=4$, the center of the $\SU(2n)$ central gauge node is $\BZ_{2n}$, but due to the screening effect of the baryons in the $D_2(\SU(2n))$ theory, it reduces to $\BZ_n$. By the same argument, this is also the case for $k=8$ and $k=12$.
\ei

Equivalently, one can simply notice from \eqref{eq:3dreduD2SU2n}-\eqref{eq:3dreduD6SU2n} that the star-shaped quivers obtained upon compactification of \eqref{eq:DpgaugingSCFTs} to 3d contain only bifundamental matter, unitary gauge groups and special unitary groups with rank $n-1$ and $2n-1$. As a result, the diagonal $\BZ_n$ subgroup of the center of the gauge groups does not act on any matter fields, leading to the conclusion that the 1-form symmetry is always at least $\BZ_n$ and enhances to $\BZ_{2n}$ whenever we do not have any $\SU(n)$ gauge nodes. This happens only for $k=6$.

\printbibliography

\end{document}